\documentclass[twoside,reqno]{bjp}
\usepackage{epsfig,cite}
\usepackage{graphics}
\usepackage{amssymb,amsmath}
\usepackage{times}
\begin{document}

\sloppy \raggedbottom

 \setcounter{page}{1}

\title {Parameter free predictions within the proxy-SU(3) model}

\runningheads{Parameter free predictions within the proxy-SU(3) model}{A. Martinou,
 D. Bonatsos, I.E. Assimakis, N. Minkov,S. Sarantopoulou, et al.}

\begin{start}
\coauthor{A. Martinou}{1},
\coauthor{D. Bonatsos}{1},
\coauthor{I.E. Assimakis}{1},
\coauthor{N. Minkov}{2},
\author{S. Sarantopoulou}{1},
\coauthor{R.B. Cakirli}{3},
\coauthor{R.F. Casten}{4,5},
\coauthor{K. Blaum}{6}

\address{Institute of Nuclear and Particle Physics, National Centre for Scientific Research ``Demokritos'', GR-15310 Aghia Paraskevi, Attiki, Greece}{1}

\address{Institute of Nuclear Research and Nuclear Energy, Bulgarian Academy of Sciences, 72 Tzarigrad Road, 1784 Sofia, Bulgaria}{2}

\address{Department of Physics, University of Istanbul, 34134 Istanbul, Turkey}{3} 

\address{Wright Laboratory, Yale University, New Haven, Connecticut 06520, USA}{4}

\address{Facility for Rare Isotope Beams, 640 South Shaw Lane, Michigan State University, East Lansing, MI 48824 USA}{5}

\address{Max-Planck-Institut f\"{u}r Kernphysik, Saupfercheckweg 1, D-69117 Heidelberg, Germany}{6}

\received{31 October 2017}

\begin{Abstract} 

Using a new approximate analytic parameter-free proxy-SU(3) scheme, we
make predictions of shape observables for deformed nuclei, namely $\beta$ and $\gamma$ deformation
variables, and compare them with empirical data and with predictions
by relativistic and non-relativistic mean-field theories. Furthermore, analytic expressions
are derived for $B(E2)$ ratios within the proxy-SU(3) model, free of
any free parameters, and/or scaling factors. The predicted
$B(E2)$ ratios are in good agreement with the experimental data for deformed
rare earth nuclides.

\end{Abstract}


\end{start}
\section{Intoduction}

The proxy-SU(3) model has been recently introduced in Refs. \cite{proxy1,proxy2}. The approximations
used in this scheme have been discussed and justified through a
Nilsson calculation in Ref. \cite{proxy1}, while in Ref. \cite{proxy2} the way to predict the $\beta$ and
$\gamma$ deformation parameters for any nucleus, using as input only the proton number
$Z$ and the neutron number $N$ of the nucleus, as well as the quantum numbers
$\lambda$ and $\mu$ appearing in the SU(3) irreducible representation (irreps) characterizing this
nucleus within the proxy-SU(3) scheme, has been described in detail.

In Section 2 of the present paper we carry out in the rare earth region a detailed
comparison of the proxy-SU(3) predictions to detailed results obtained with the
D1S Gogny interaction, tabulated in Ref. \cite{Gogny}, while in Section 3 we calculate
$B(E2)$ ratios within ground state bands and $\gamma_1$ bands of some deformed rare
earth nuclei and compare them to the existing data \cite{ENSDF}. In both sections, no free parameters
are used.

\section{Predictions for the deformation parameters}

\subsection{Connection between deformation variables and SU(3) quantum numbers}

A connection between the collective variables $\beta$ and $\gamma$ of the collective model \cite{BM}
and the quantum numbers $\lambda$ and $\mu$ characterizing the irreducible representation 
$(\lambda,\mu)$ of SU(3) \cite{Elliott1,Elliott2} has long been established 
\cite{Castanos,Park}, based on the fact that the invariant quantities of the two theories should possess the same values. 

The relevant equation for $\beta$ reads 
\cite{Castanos,Park}
\begin{equation}\label{b1}
	\beta^2= {4\pi \over 5} {1\over (A \bar{r^2})^2} (\lambda^2+\lambda \mu + \mu^2+ 3\lambda +3 \mu +3), 
\end{equation}
where $A$ is the mass number of the nucleus and $\bar{r^2}$ is related to the dimensionless mean square radius \cite{Ring}, $\sqrt{\bar{r^2}}= r_0 A^{1/6}$. 
The constant $r_0$ is determined from a fit over a wide range of nuclei \cite{DeVries,Stone}. We use the value in Ref. \cite{Castanos}, $r_0=0.87$, in agreement to Ref. \cite{Stone}.
The quantity in Eq. (1) is proportional to the second order Casimir operator of SU(3) \cite{IA}, 
 \begin{equation}\label{C2} 
 C_2(\lambda,\mu)= {2 \over 3} (\lambda^2+\lambda \mu + \mu^2+ 3\lambda +3 \mu). 
\end{equation}

The relevant equation for $\gamma$ reads \cite{Castanos,Park}
\begin{equation}\label{g1}
\gamma = \arctan \left( {\sqrt{3} (\mu+1) \over 2\lambda+\mu+3}  \right). 
\end{equation}

\subsection{Numerical results}\label{bet}

In Fig. 1 (Fig. 2) results for the collective variable $\beta$ ($\gamma$) are shown, calculated from Eq. (\ref{b1}) [Eq. (\ref{g1})] and rescaled in the case of $\beta$ as described in detail in Ref. \cite{proxy2}. Experimental results obtained
from Ref. \cite{Raman} are also shown for comparison, as in Ref. \cite{proxy2}. Furthermore,
comparison to the detailed results provided by the D1S Gogny force, tabulated
in Ref. \cite{Gogny}, is made. By ``Gogny D1S mean'' we label the mean ground state 
$\beta$ ($\gamma$) deformation [entry 11 (12) in the tables of \cite{Gogny}], while the error bars correspond to the variance of the ground state $\beta$ ($\gamma$) deformation [entry 13 (14) in \cite{Gogny}]. By ``Gogny D1S min.'' we label the $\beta$ ($\gamma$) deformation at the HFB energy minimum [entry 4 (5) in \cite{Gogny}]. In the case of $\beta$, predictions obtained with relativistic mean field theory (RMF) \cite{Lalazissis} are also shown.

In the case of $\beta$ we note that the HFB minimum lies always within the error bars of the D1S Gogny mean g.s. deformation (except for $N = 84$), while in the case of $\gamma$ we see
that the HFB minimum lies well below the error bars of the D1S Gogny mean g.s.
deformation for most of the $N$ values, but jumps suddenly to very high values,
close to 60 degrees, at or near $N = 116$.

In Fig. 1 the proxy-SU(3) predictions for $\beta$ lie within the error bars of the D1S Gogny mean g.s. deformation, with the following few exceptions: a) The first ($N = 84$) point
in Gd-Hf, b) the last two ($N = 120$, 122) points in Gd and Dy, as well as the last
point ($N = 122$) in Er, c) a few isolated cases, like the $N = 110$ point in Er, the
$N = 102$ point in W and Os, and the $N = 100$, 102 points in Pt. We stress, however, that proxy-SU(3) is only valid for deformed nuclei and therefore some of these differences [items a) and b)] may not be meaningful.

Similar observations can be made for $\gamma$ in Fig. 2, where the exceptions occur in:
a) The first three points ($N =84$, 86, 88) in Gd-Yb, b) the last three ($N = 118$, 120, 122) points in W-Pt, which agree with the HFB minimum rather than with the mean g.s. deformation,
c) a few isolated cases, like some of the $N = 106$ in Gd, Dy, Er, Hf, W, and
several points in Yb. 

\subsection{Discussion}

The above observations can be summarized as follows:

1) While the $\beta$ deformation at the HFB energy minumum remains always close
to the mean ground state $\beta$ deformation, the behavior of the $\gamma$ deformation is
strikingly different. In most of the region the $\gamma$  deformation at the HFB energy
minumum remains close to zero, but it suddenly jumps to values close to 60
degrees near the end of the shell ($N = 116$-122). This jump is sudden in Gd-Hf,
while it becomes more gradual in W, Os, Pt.

2) In the beginning of the region we see some failures of proxy-SU(3) at $N = 84$,
86, 88 in Gd-Hf. These failures are expected, since these nuclei are not well
deformed, as known from their $R_{4/2}$ ratios.

3) In most of the region, the proxy-SU(3) predictions for both $\beta$ and $\gamma$
are in good agreement with the D1S Gogny mean g.s. deformations.

4) The agreement of the proxy-SU(3) predictions with the D1S Gogny mean g.s.
deformations remains good up to the end of the shell for $\beta$, while for $\gamma$ in W, Os, Pt it is observed that the proxy-SU(3) predictions for $\gamma$ jump at the end of the shell from close agreement to the D1S Gogny mean g.s. deformations to close
agreement with $\gamma$ at the HFB energy minimum, i.e., close to 60 degrees.

\section{B(E2) ratios}

As discussed in Appendix A, $B(E2)$s within the proxy-SU(3) model are proportional
to the square of the relevant reduced matrix element of the quadrupole
operator $Q$. If ratios of $B(E2)$s within the same nucleus and within the same irreducible representation are considered, only
the relevant SU(3)$\to$SO(3) coupling coefficients remain, while all other factors
cancel out, leading to
\begin{equation} 
{B(E2;L_i\to L_f)\over B(E2;2_g\to 0_g)} = 5 {2L_f+1\over 2L_i+1} \\
{( \langle (\lambda,\mu) K_i L_i ; (1,1)2 || (\lambda,\mu) K_f L_f \rangle)^2 \over 
( \langle (\lambda,\mu) 0 2 ; (1,1)2 || (\lambda,\mu) 0 0 \rangle)^2}, \label{BE2}
\end{equation} 
where normalization to the $B(E2)$ connecting the first excited $2^+$ state to the
$0^+$ ground state of even-even nuclei is made. The needed SU(3)$\to$SO(3)
coupling coefficients are readily obtained from the SU3CGVCS code \cite{RBcode}, as
described in Appendix A.

It should be noticed that the ratios given by Eq. (\ref{BE2}) are completely free of any
free parameters and/or scaling factors.

\subsection{ Numerical results}

Calculations have been performed for the proxy-SU(3) irreps (54,12), (52,14),
and (50,10). The irrep (54,12) accommodates $^{168}$Er, for which complete spectroscopy
has been performed \cite{Warner}, and $^{160}$Gd, for which little data on $B(E2)$s
exist \cite{ENSDF}. The irrep (52,14) accommodates $^{162}$Dy, for which complete spectroscopy has been performed \cite{Aprahamian}, and $^{166}$Er, for which rich 
data exist \cite{ENSDF}. It also
accommodates $^{172}$Er, for which little data on $B(E2)$s exist \cite{ENSDF}. The irrep (50,10)
accommodates $^{156}$Gd, which has been cited as the textbook example of the bosonic SU(3)
in the IBM-1 framework \cite{IA}. The Alaga values \cite{Casten}, derived from the relevant Clebsch-Gordan coefficients alone, are also given for comparison. 

\subsection{ Comparisons to experimental data for specific nuclei}\label{Exp}

$B(E2)$s within the ground state band are shown in Fig. 3. Agreement between the
proxy-SU(3) predictions and the data is excellent in the cases of $^{156}$Gd, 
$^{162}$Dy, and $^{166}$Er, while in $^{168}$Er three points are missed. 
It appears that nuclear stretching \cite{Pin} has been properly taken into account.

In Fig. 4 three pairs of nuclei, each pair accommodated within a single proxy-
SU(3) irrep, are shown. These are the only pairs for which adequate data exist \cite{ENSDF} in
the region of 50-82 protons and 82-126 neutrons. Agreement within the experimental
errors is seen in almost all cases.

Proxy-SU(3) predictions for $B(E2)$s within the $\gamma_1$ band, with $\Delta L=-2$ (increasing
with $L$) and $\Delta L=-1$ (decreasing with $L$), are shown in Fig. 5, and
are compared to the data for nuclei for which sufficient data exist \cite{ENSDF}. The distinction
between increasing $B(E2)$s with $\Delta L = -2$ and decreasing $B(E2)$s with 
$\Delta L = -1$ is seen clearly in the data.

\subsection{ Discussion}

The main findings of the present section can be summarized as follows.

Analytic expressions for $B(E2)$ ratios for heavy deformed nuclei providing
numerical results in good agreement with experiment are derived within
the proxy-SU(3) scheme without using any free parameters
and/or scaling factors. The derivation, described in Appendix A, is exact.  The only quantities appearing in the final formula are the relevant
SU(3)$\to$SO(3) coupling coefficients, for which computer codes are readily available
\cite{RBcode,Akicode}.

Concerning further work, spectra of heavy deformed nuclei will be considered
within the proxy-SU(3) scheme, involving three- and/or four-body terms in order
to break the degeneracy between the ground state and $\gamma_1$ bands  \cite{DW1,DW2,PVI}. Furthermore,
$B(M1)$ transition rates can be considered along the proxy-SU(3) path, using
the techniques already developed \cite{CDL} in the framework of the pseudo-SU(3)
scheme.

\section*{Acknowledgements}

Work partly supported by the Bulgarian National Science Fund (BNSF) under Contract No. DFNI-E02/6, by the US DOE under Grant No. DE-FG02- 91ER-40609, and by the MSU-FRIB laboratory, by the Max Planck Partner group, TUBA-GEBIP, and by the Istanbul University Scientific Research Project No. 54135.

\section*{Appendix A. Formulae used for $B(E2)$s}

In most of the earlier work, effective charges
\begin{equation}
e_\pi=e+e_{eff}, \qquad e_\nu=e_{eff}, 
\end{equation}
have been used, where the effective charge $e_{eff}$ is usually fixed so that the calculated
$B(E2)$ transition rate for the $2_1^+\to 0_1^+$ transition reproduces the experimental
value \cite{CDL}. In the present approach we make the choice $e_{eff}=0$, which leads to 
$e_\pi = e$ and $e_\nu = 0$.

The needed matrix elements of the relevant quadrupole operators, $Q^\pi$ and $Q^\nu$ for
protons and neutrons respectively, are given in detail in Appendix D of Ref. \cite{DW2},
with 
SU(3)$\to$SO(3) coupling coefficients \cite{RBcode,Akicode,Akiyama,RB}, as well as 9-$(\lambda,\mu)$ coefficients \cite{Akicode,Akiyama,Millener}
appearing in the relevant expressions. Codes for calculating
these coefficients are readily available, given in the references just cited. With
$e_{eff} = 0$ one sees that only the matrix elements of $Q^\pi$ are needed. Furthermore,
if we use ratios of $B(E2)$ transition rates within a given nucleus, 
the 9-$(\lambda,\mu)$
coefficients will cancel out and the only nontrivial term remaining in the B(E2)
ratios will be the ratio of the relevant SU(3)$\to$SO(3) coupling coefficients, which
remarkably involve only the highest weight $(\lambda,\mu)$ irrep characterizing the whole nucleus, while
they are independent of the $(\lambda_\pi,\mu_\pi)$ and $(\lambda_\nu,\mu_\nu)$ irreps characterizing the protons
and the neutrons separately.


\begin{figure}[b]
\epsfig{file=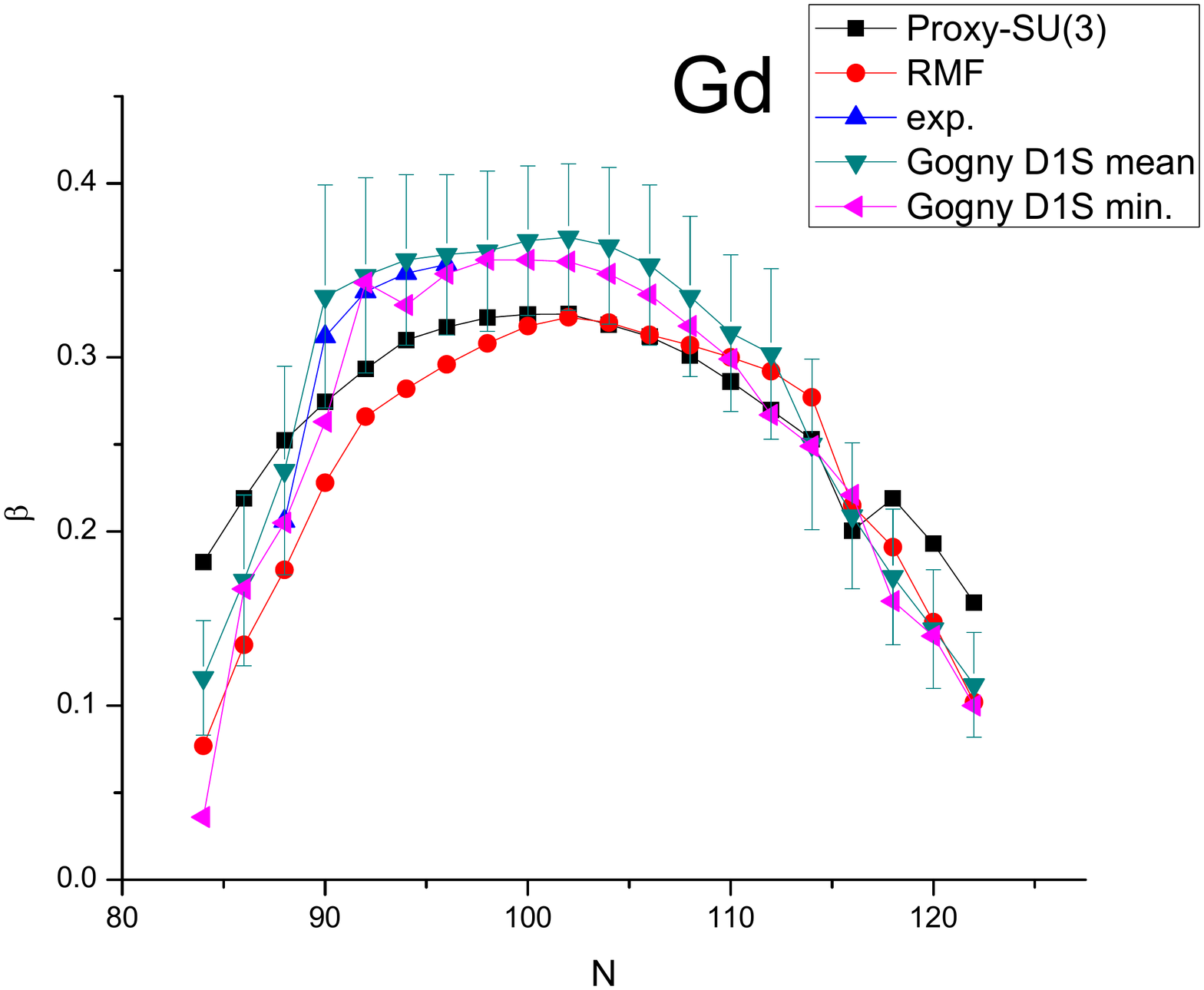,width=55mm}
\epsfig{file=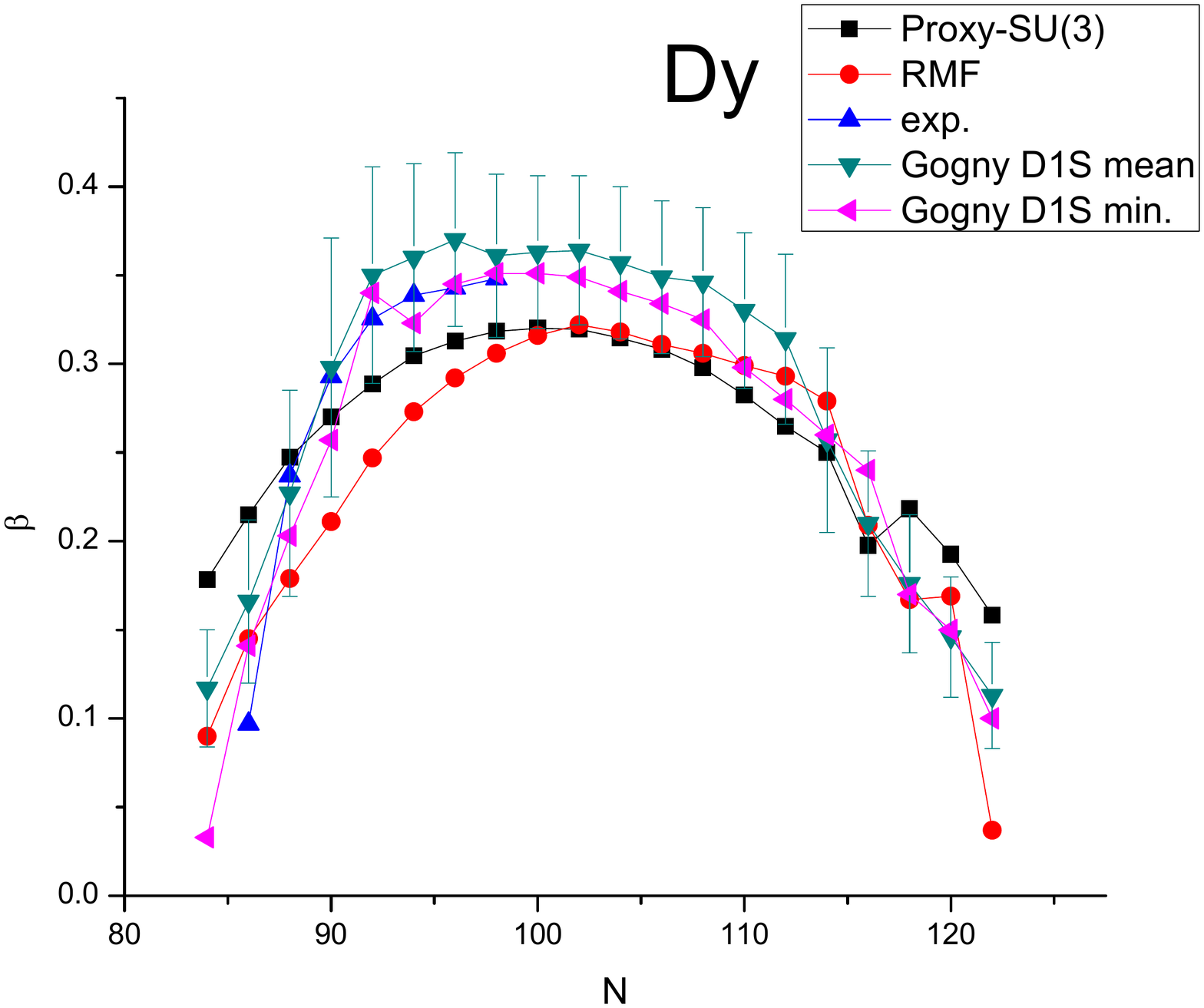,width=55mm}

\epsfig{file=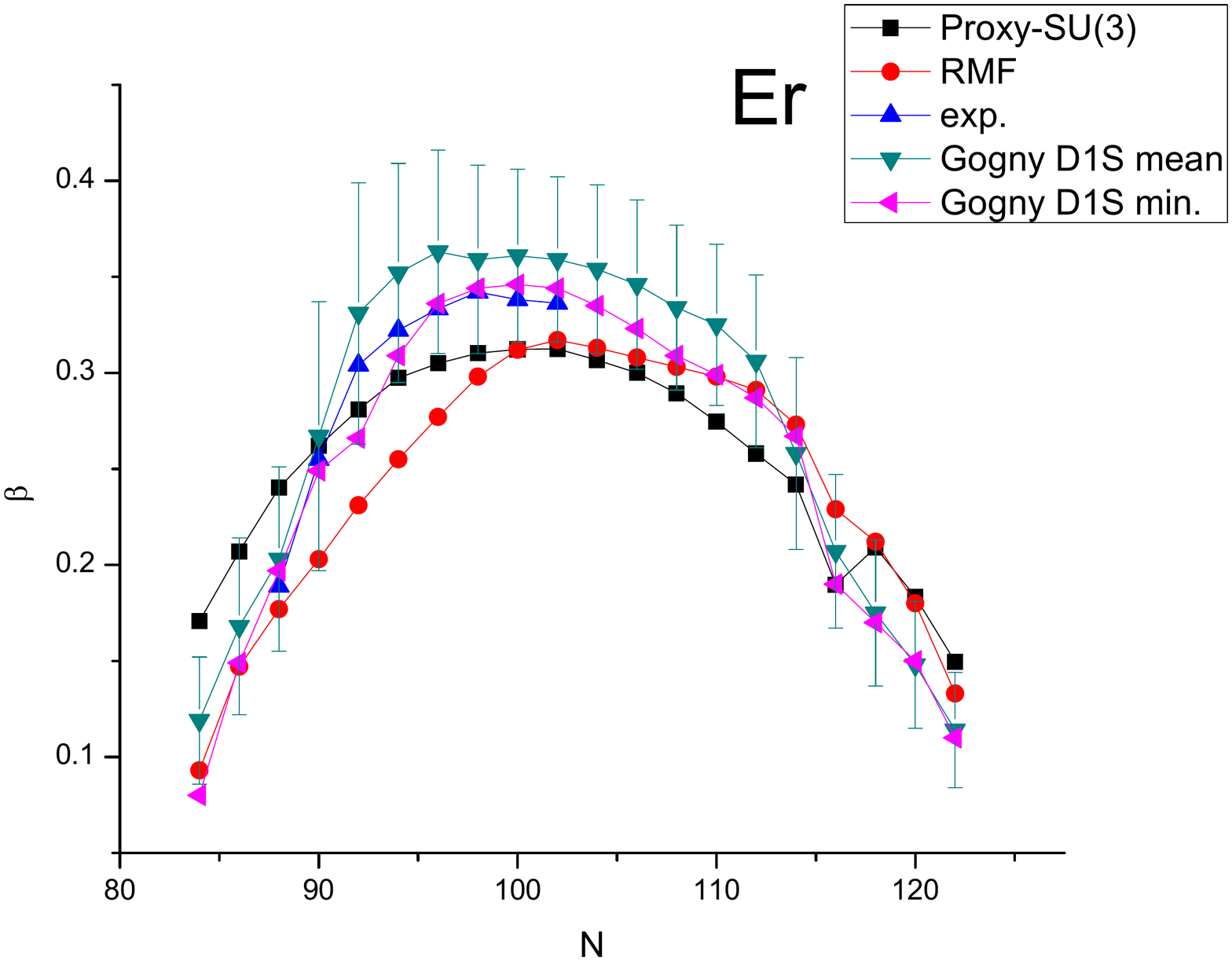,width=55mm}
\epsfig{file=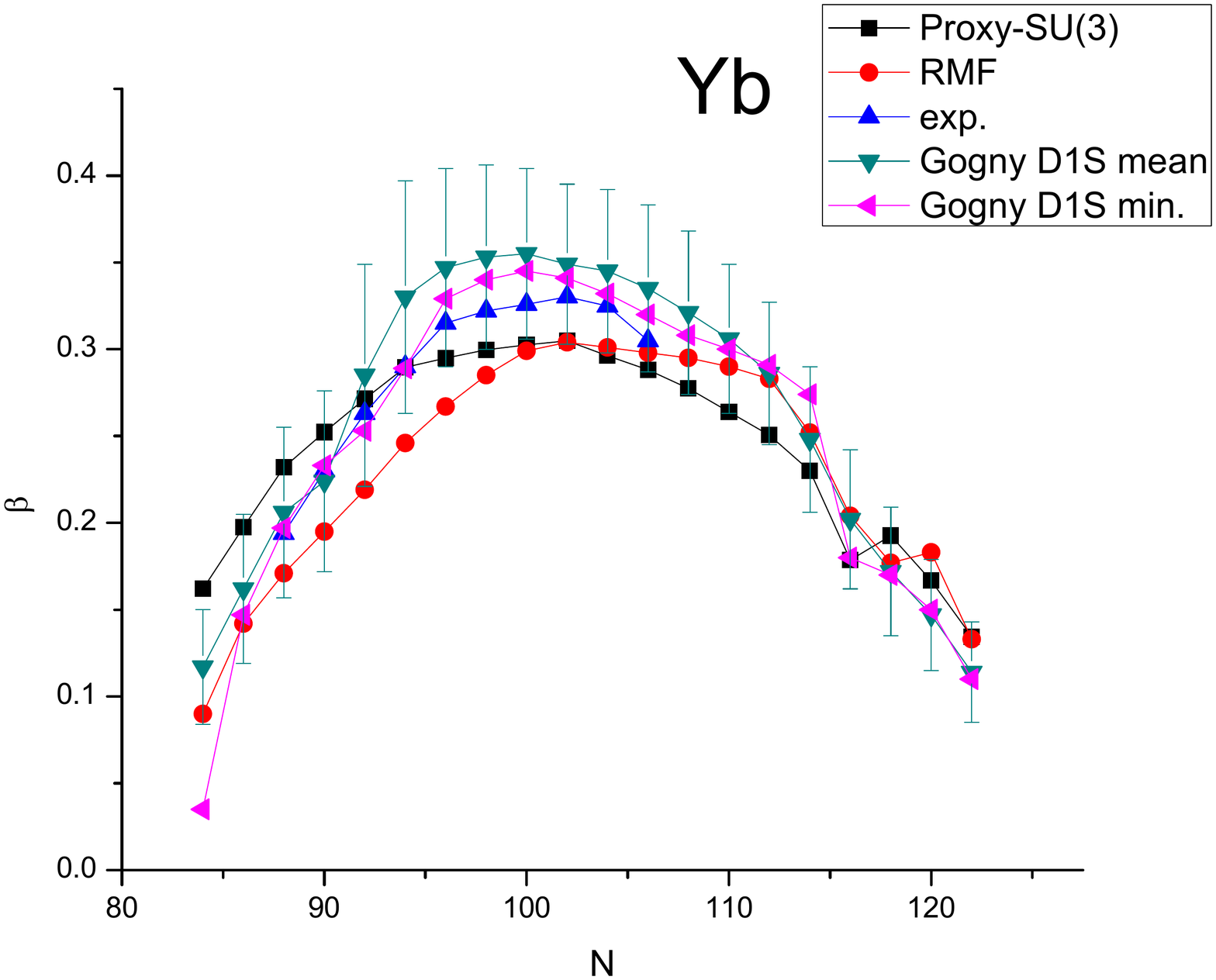,width=55mm}

\epsfig{file=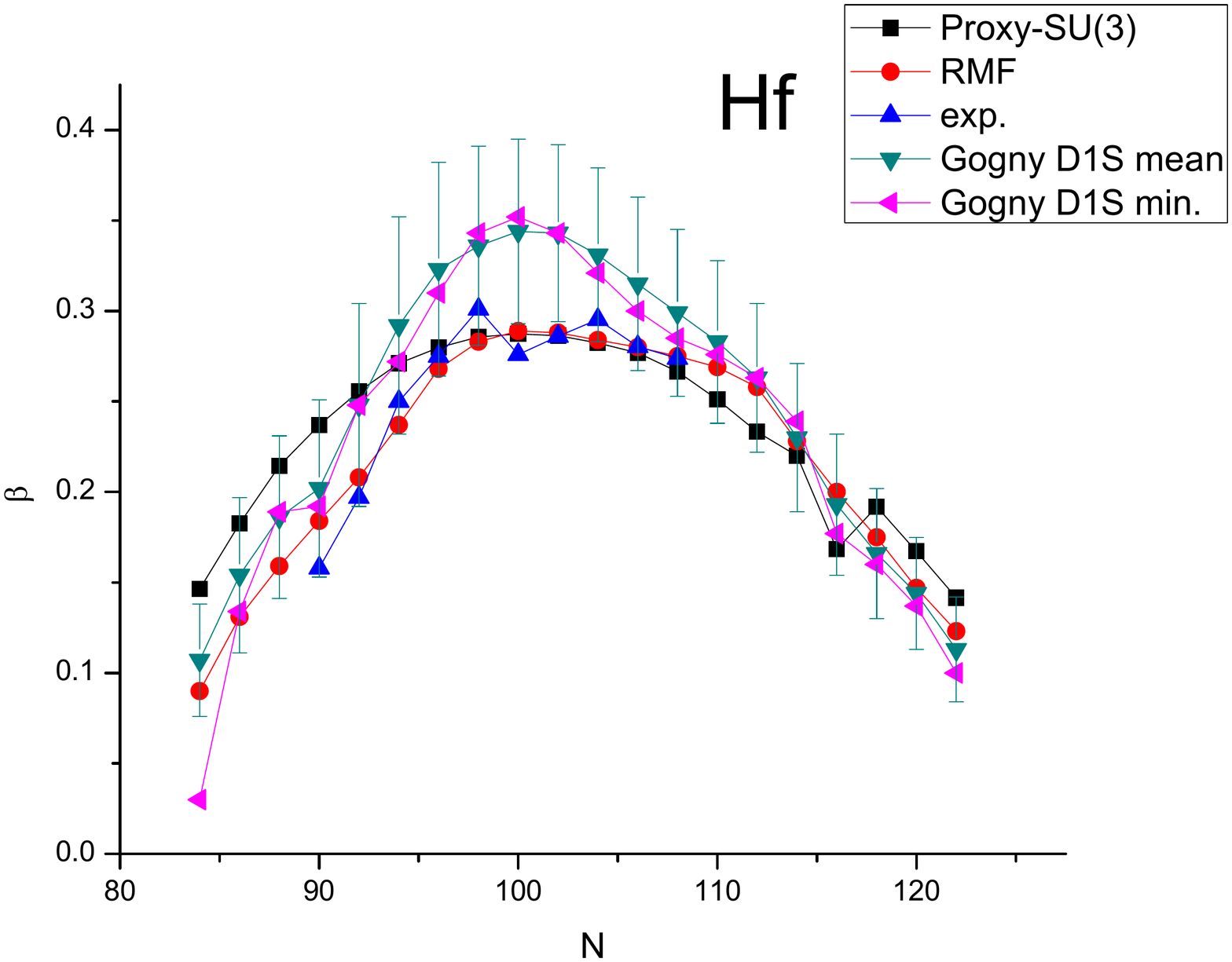,width=55mm}
\epsfig{file=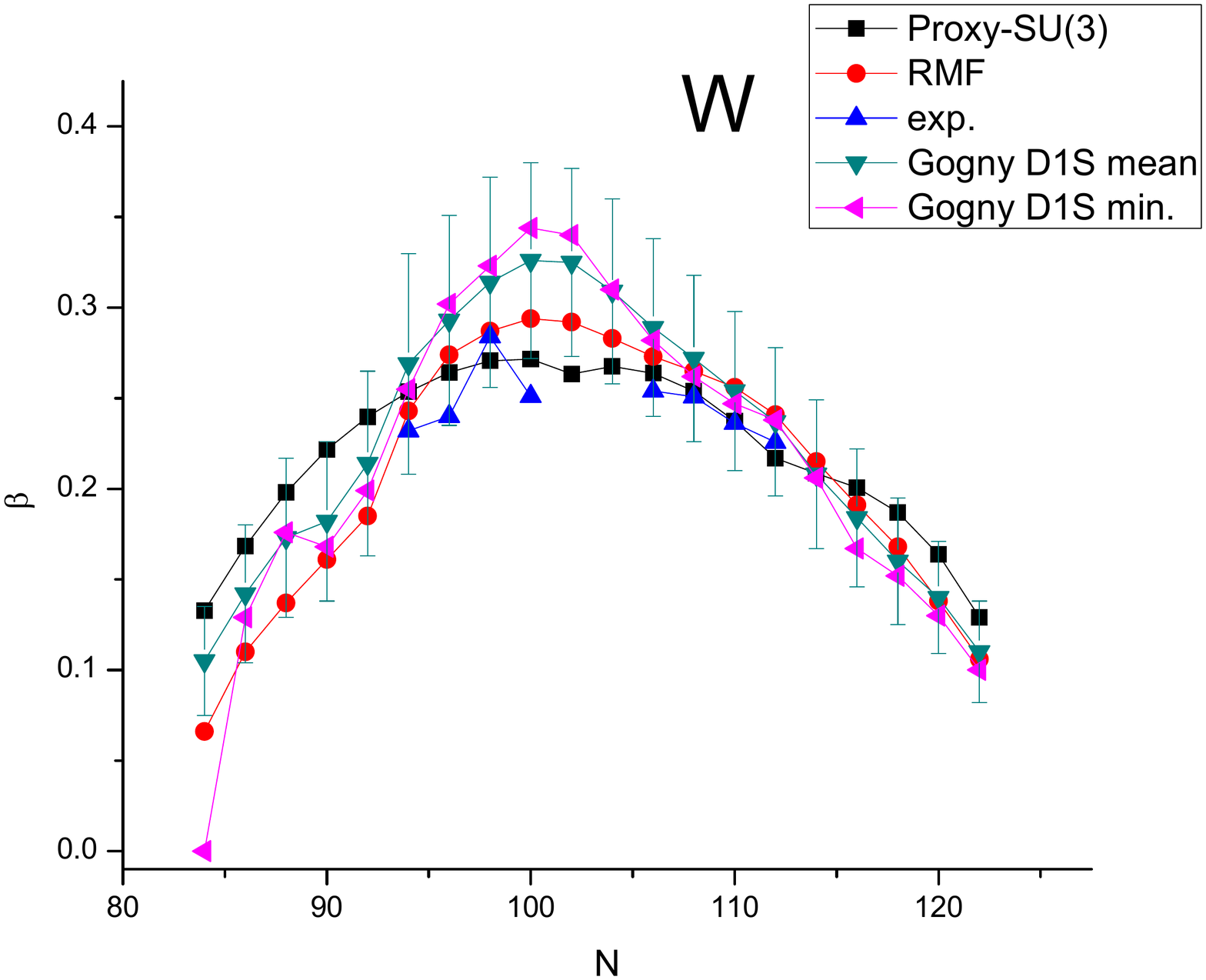,width=55mm}

\epsfig{file=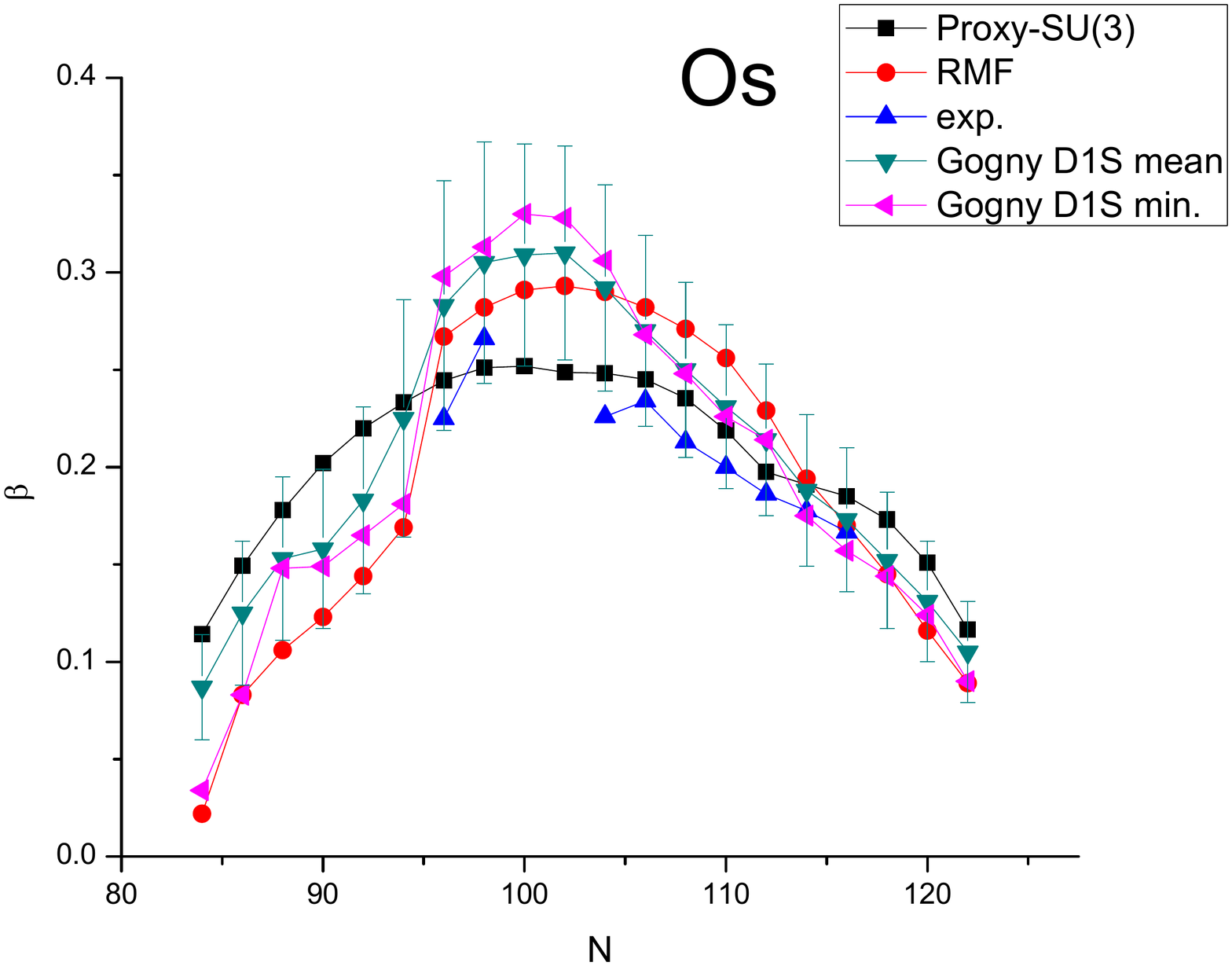,width=55mm}
\epsfig{file=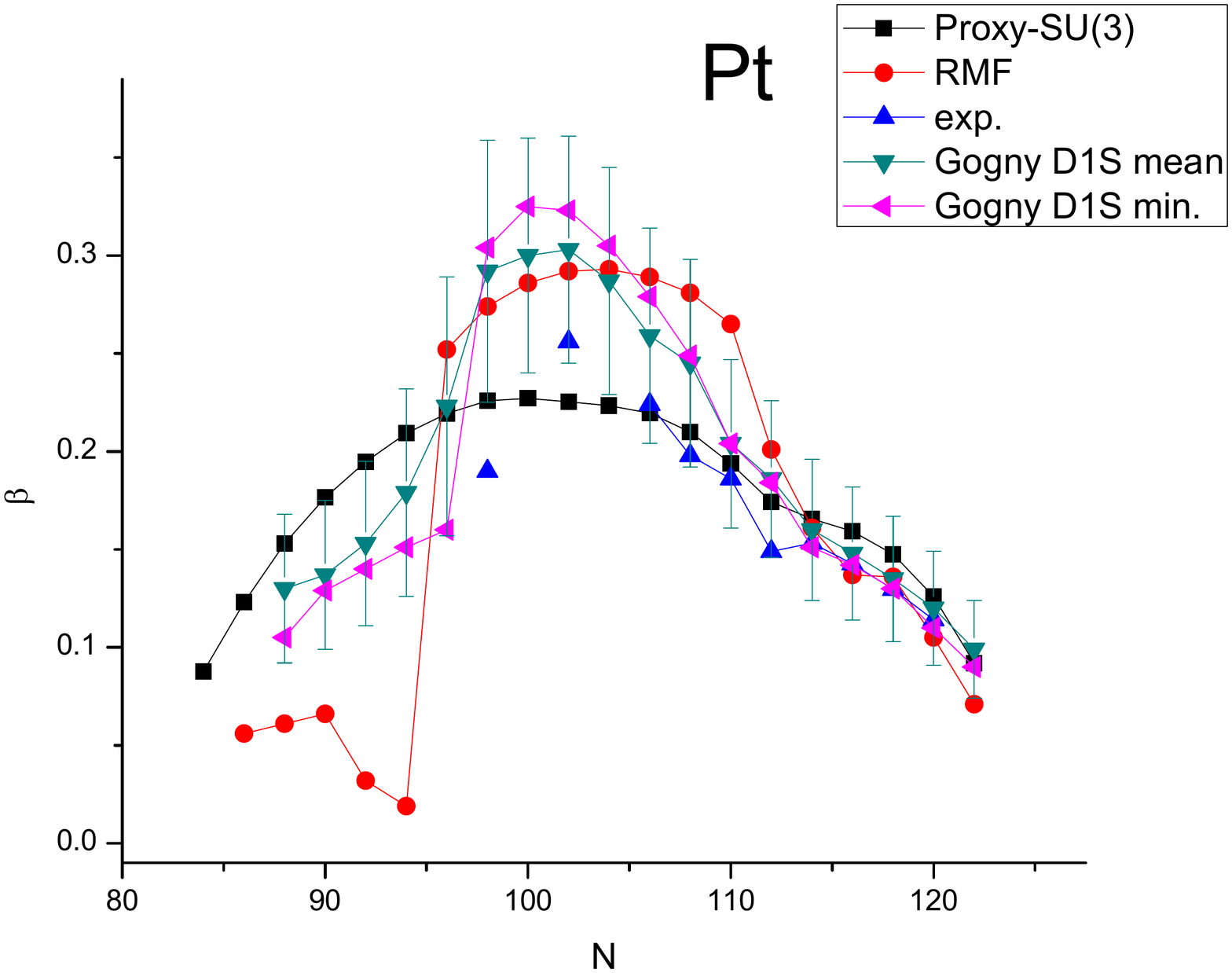,width=55mm}

\caption{Proxy SU(3) predictions for Gd-Pt isotopes  for $\beta$, obtained from Eq.  
 (\ref{b1}), as described in detail in Ref.
\cite{proxy2}, compared with results by the D1S-Gogny interaction (D1S-Gogny) 
\cite{Gogny} and by relativistic mean field theory (RMF) \cite{Lalazissis}, 
as well as with empirical values (exp.) \cite{Raman}. 
 See subsection \ref{bet} for further discussion.} 

\end{figure}


\begin{figure}[b]
\epsfig{file=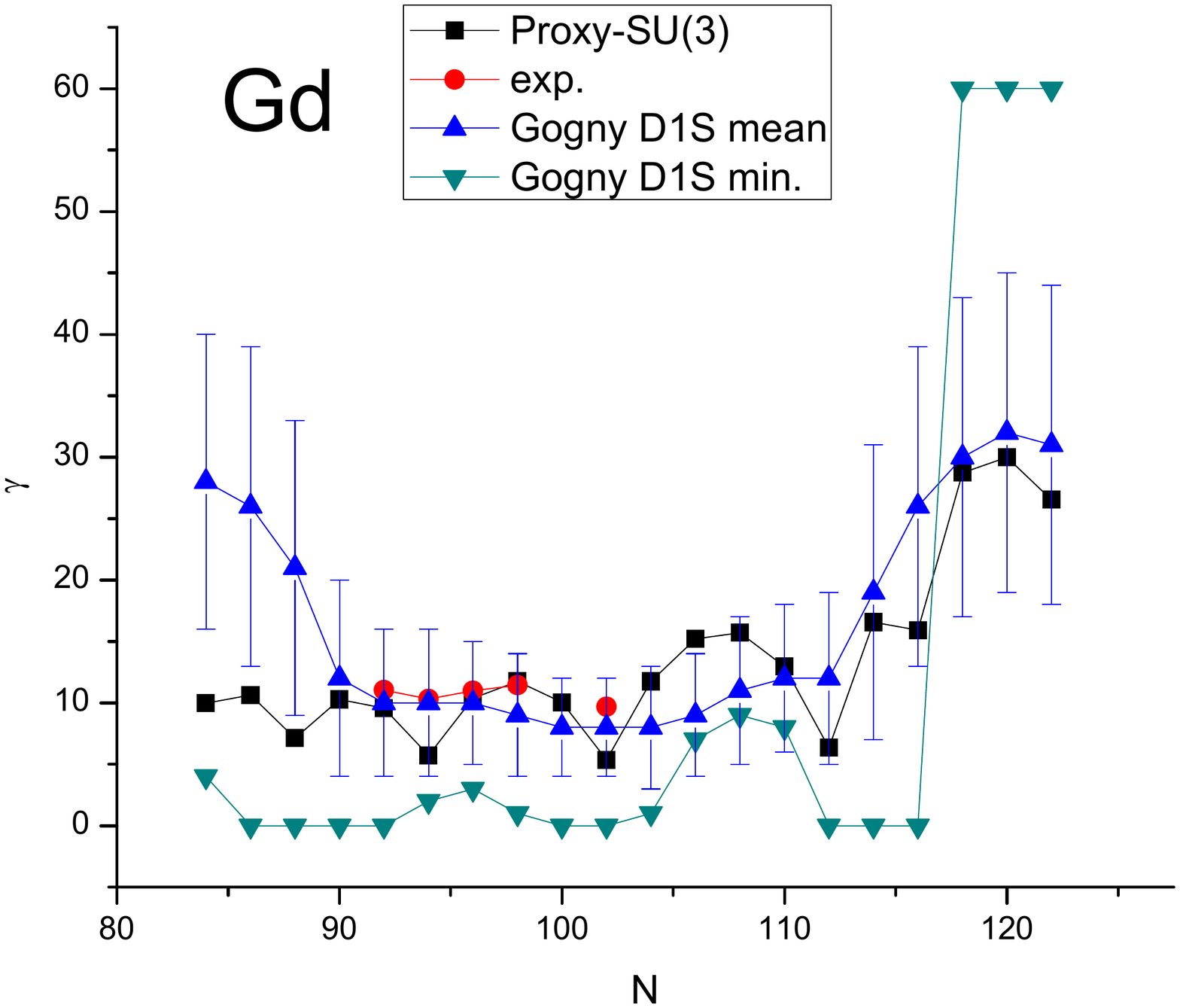,width=55mm}
\epsfig{file=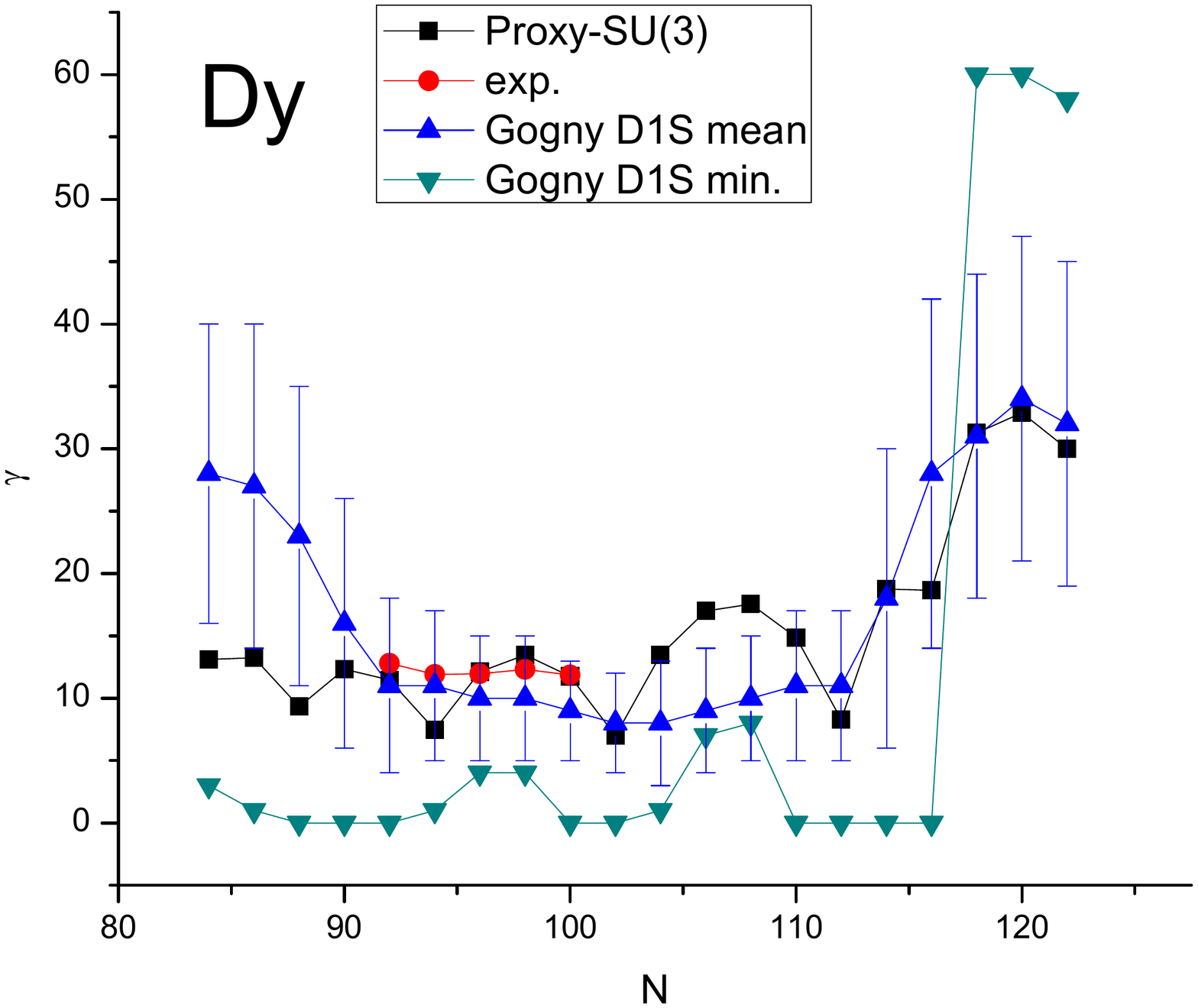,width=55mm}

\epsfig{file=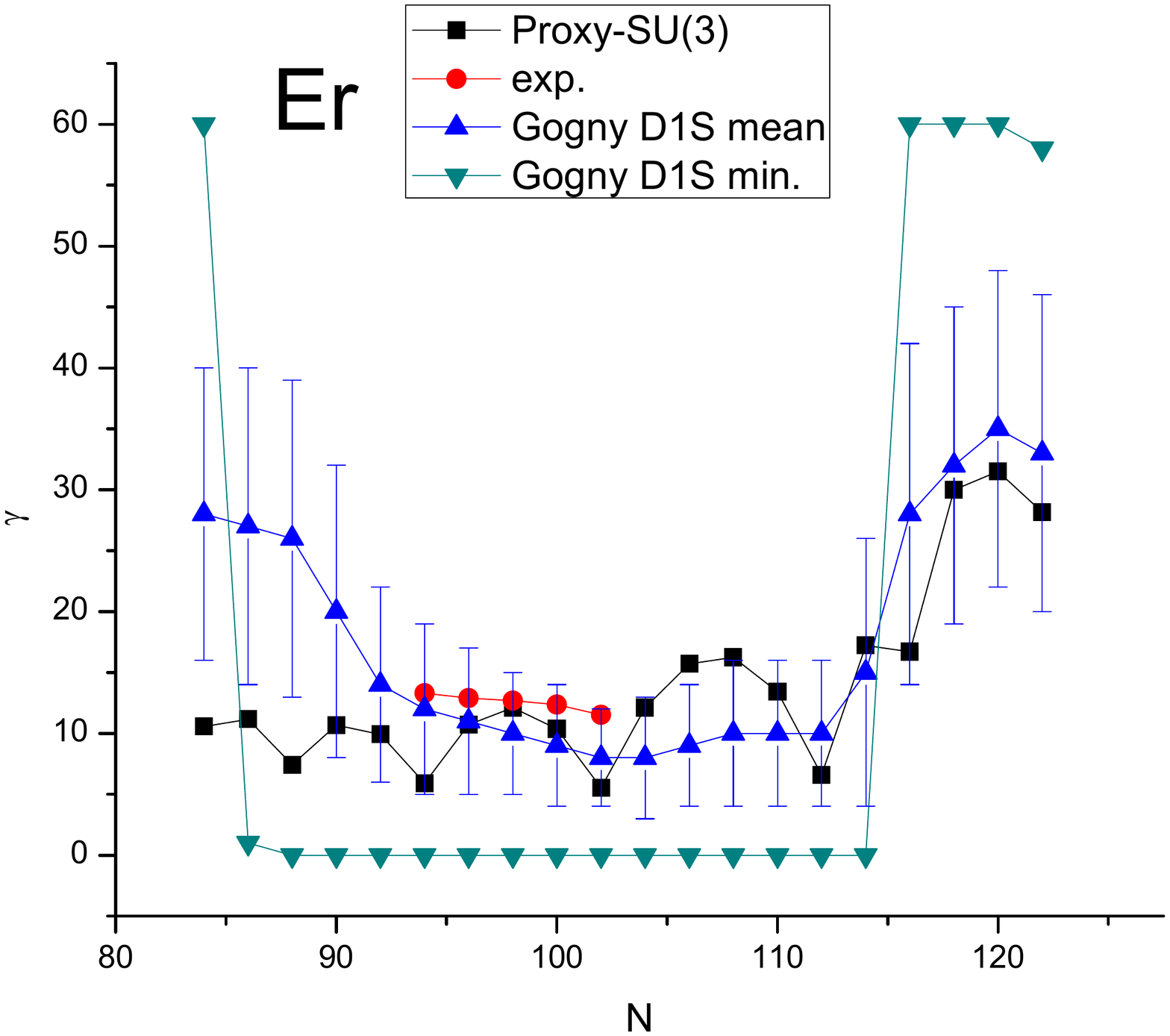,width=55mm}
\epsfig{file=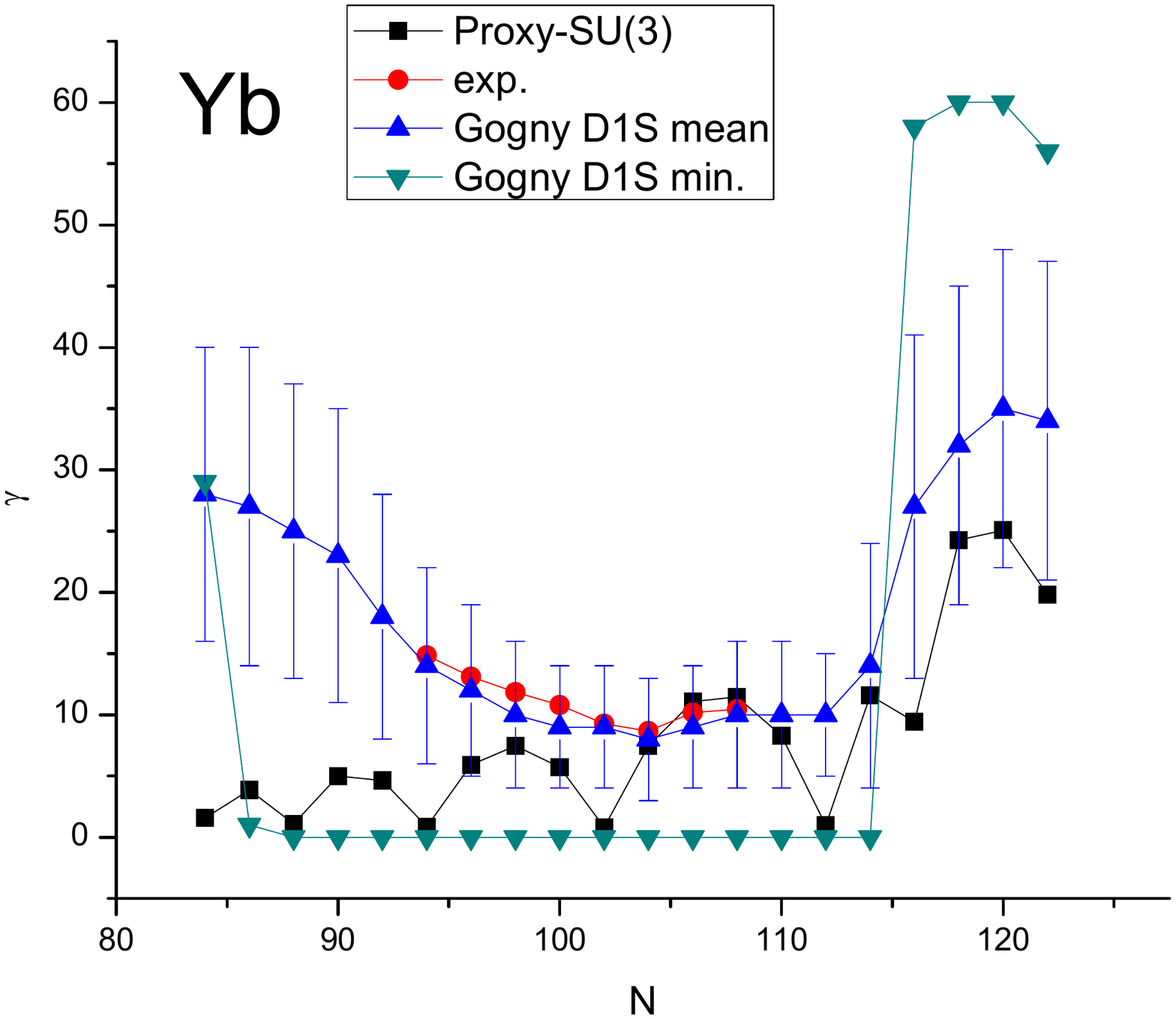,width=55mm}

\epsfig{file=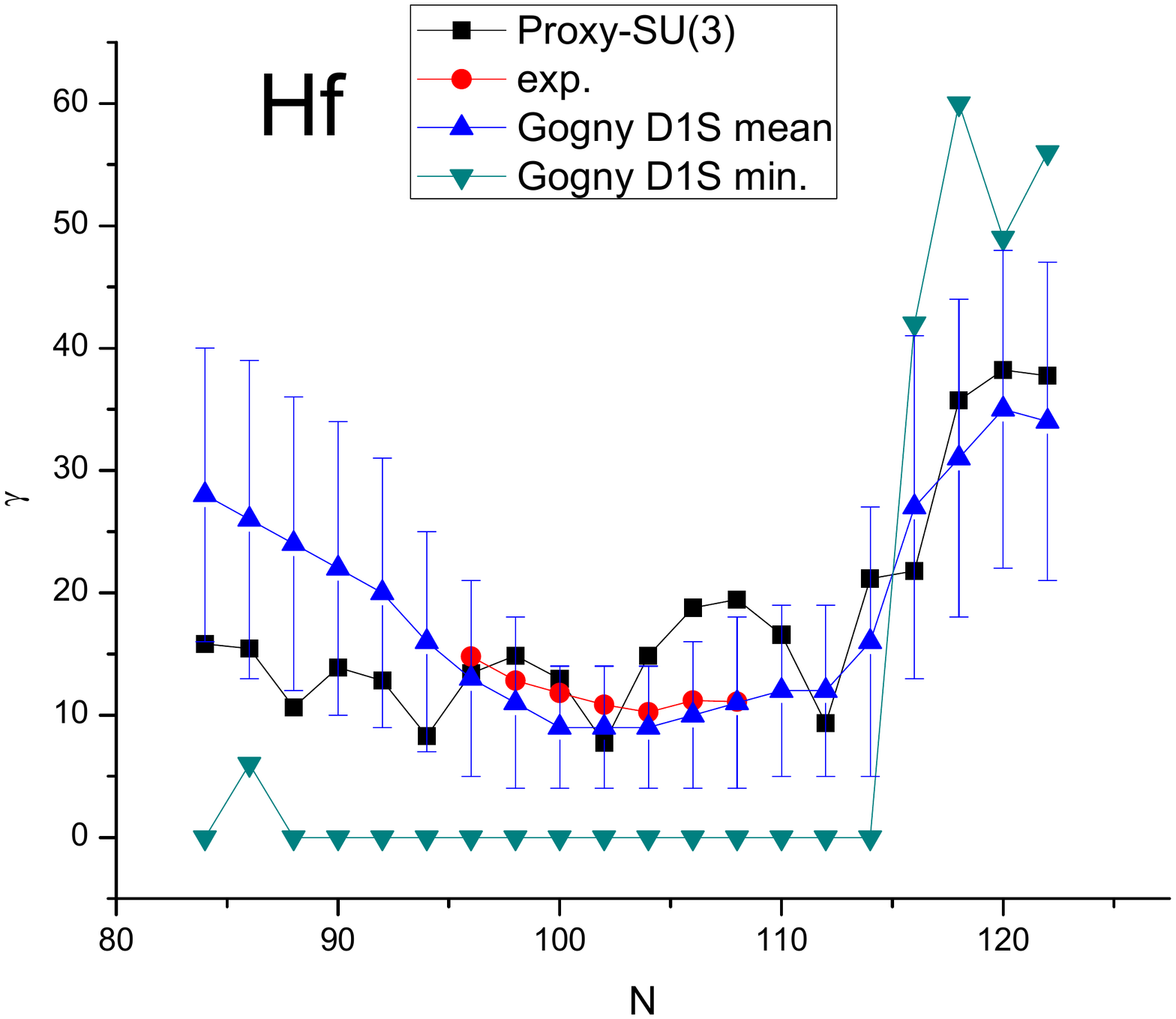,width=55mm}
\epsfig{file=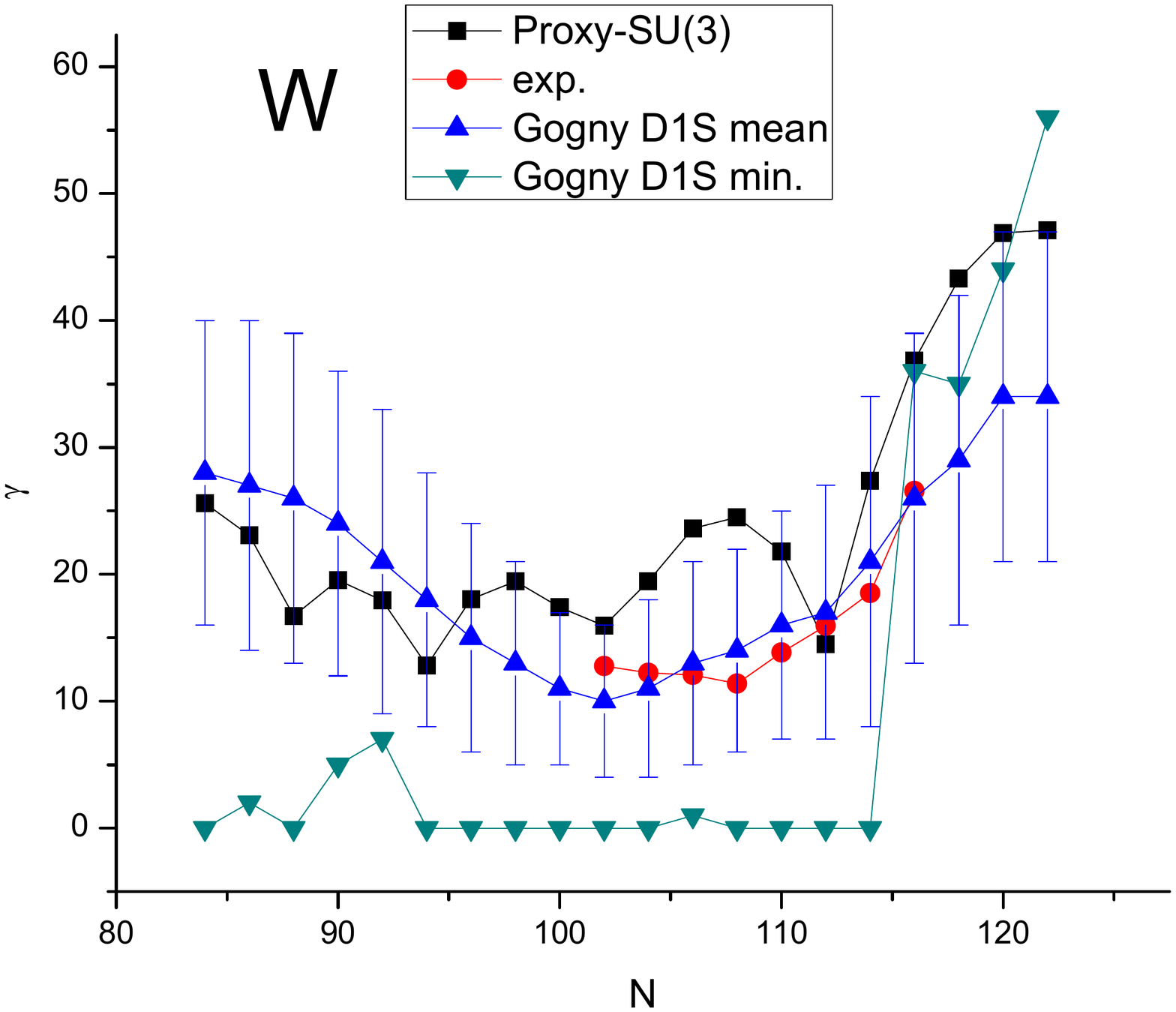,width=55mm}

\epsfig{file=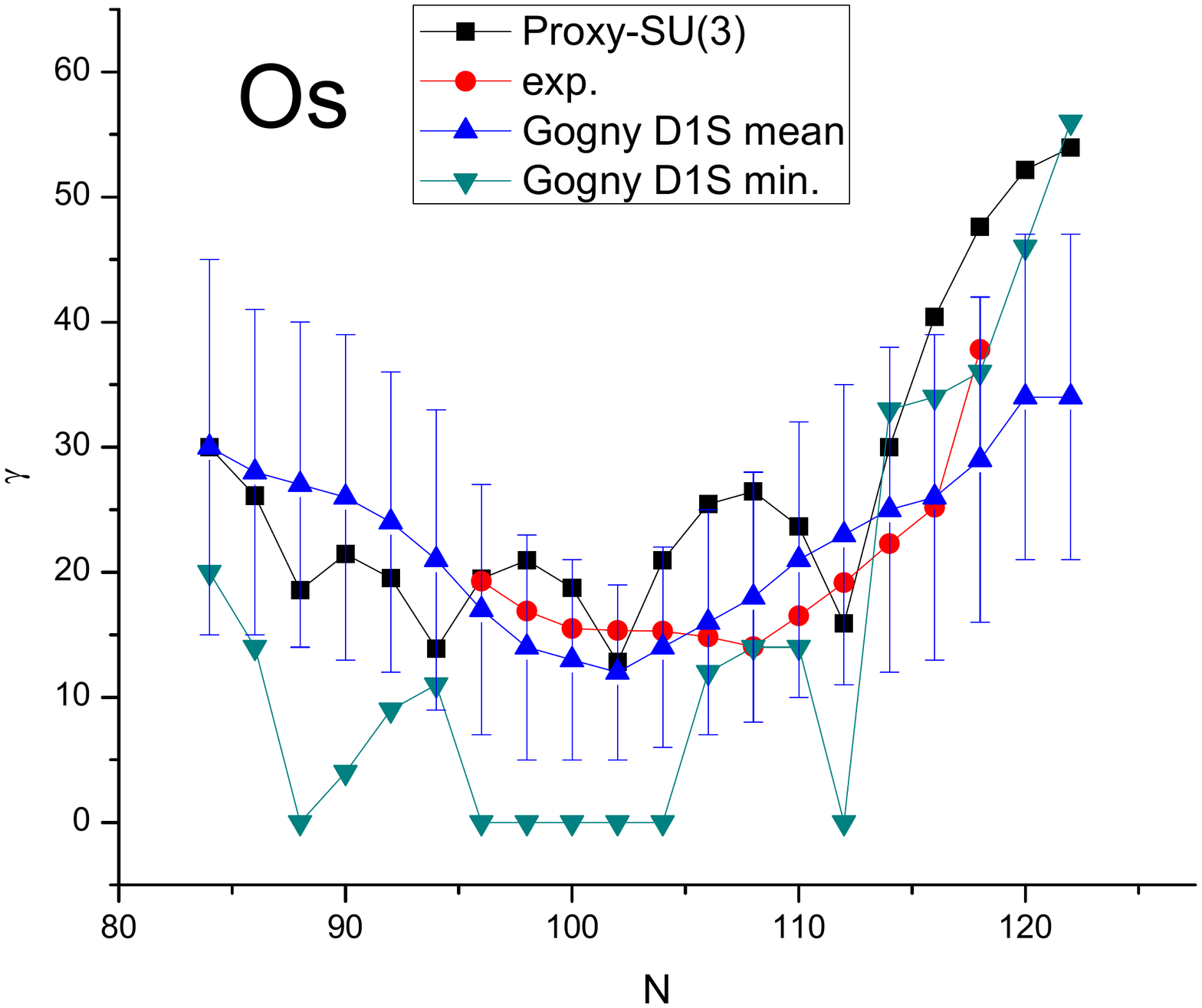,width=55mm}
\epsfig{file=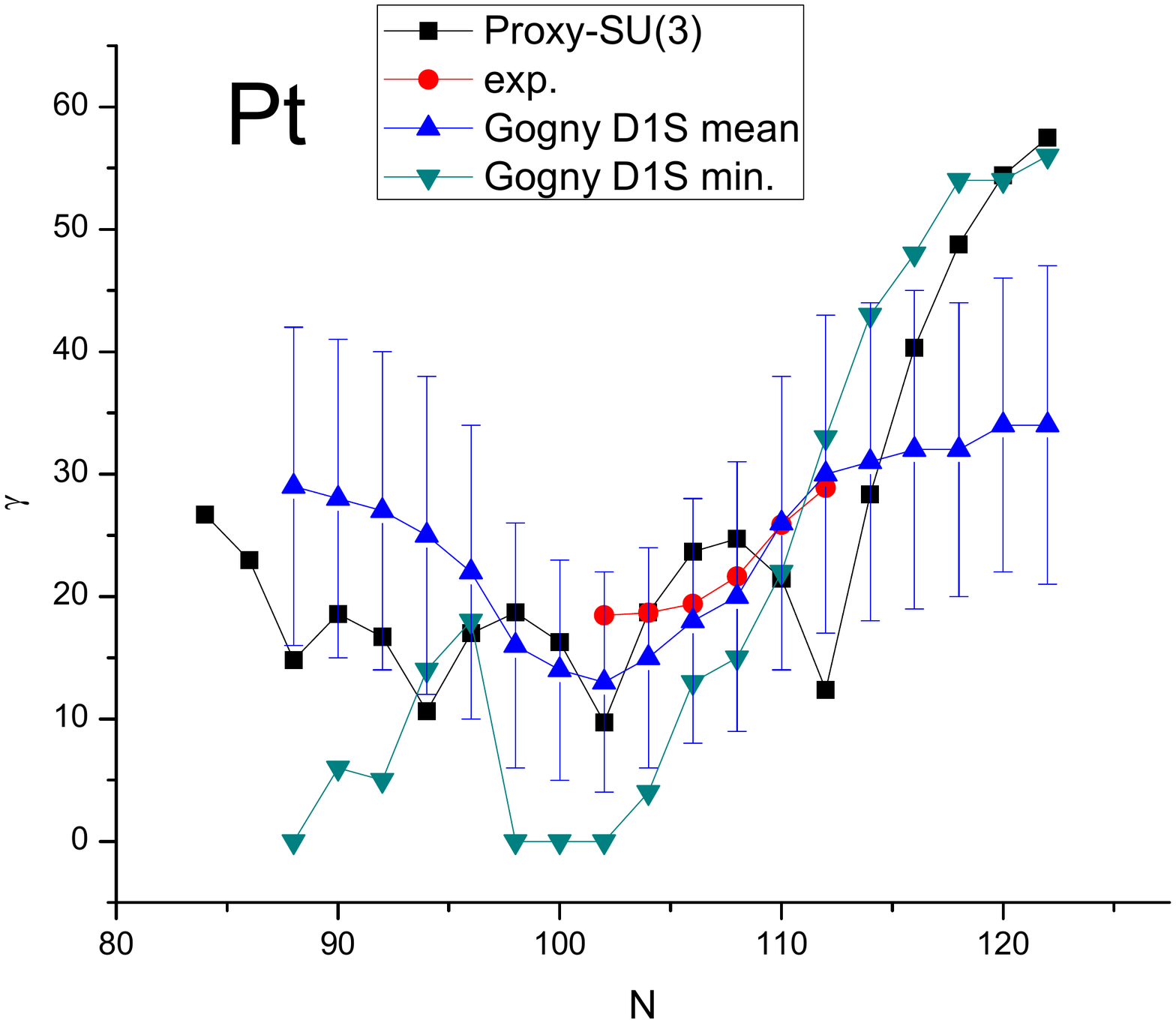,width=55mm}

\caption{Same as Fig. 1, but for $\gamma$, derived from Eq. (\ref{g1}). See subsection \ref{bet} for further discussion.}

\end{figure}


\begin{figure}[b]

\epsfig{file=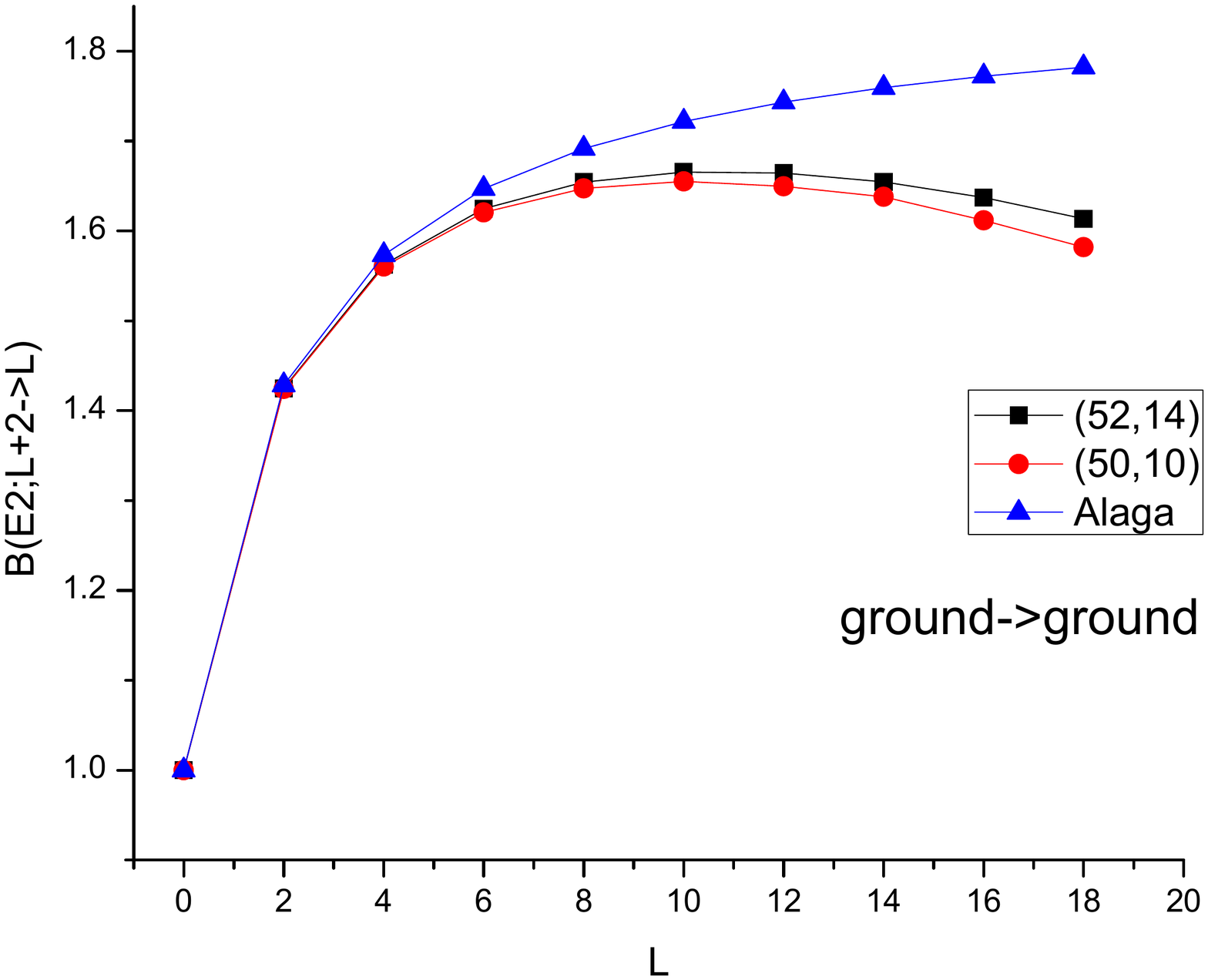,width=55mm}
\epsfig{file=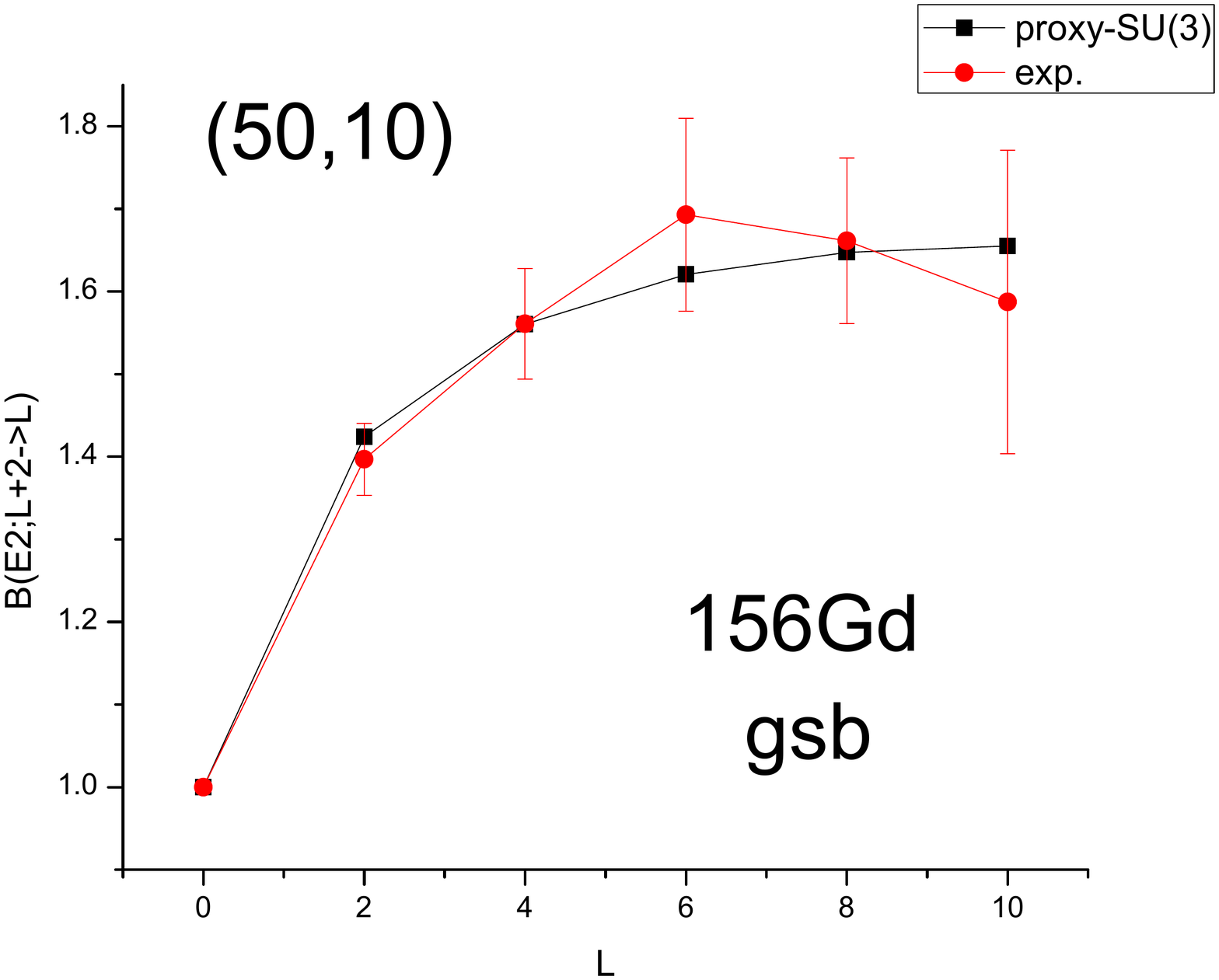,width=55mm}

\epsfig{file=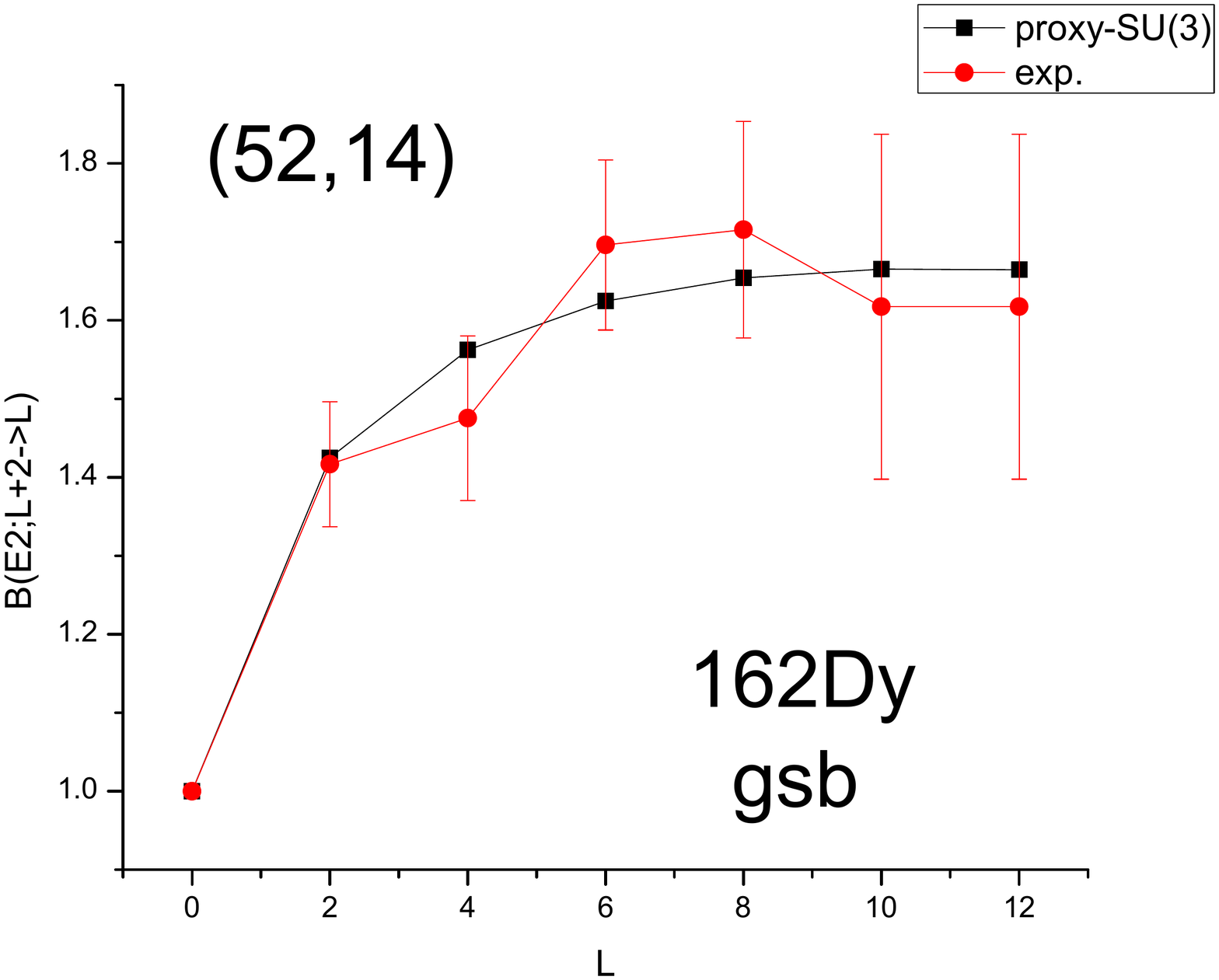,width=55mm}
\epsfig{file=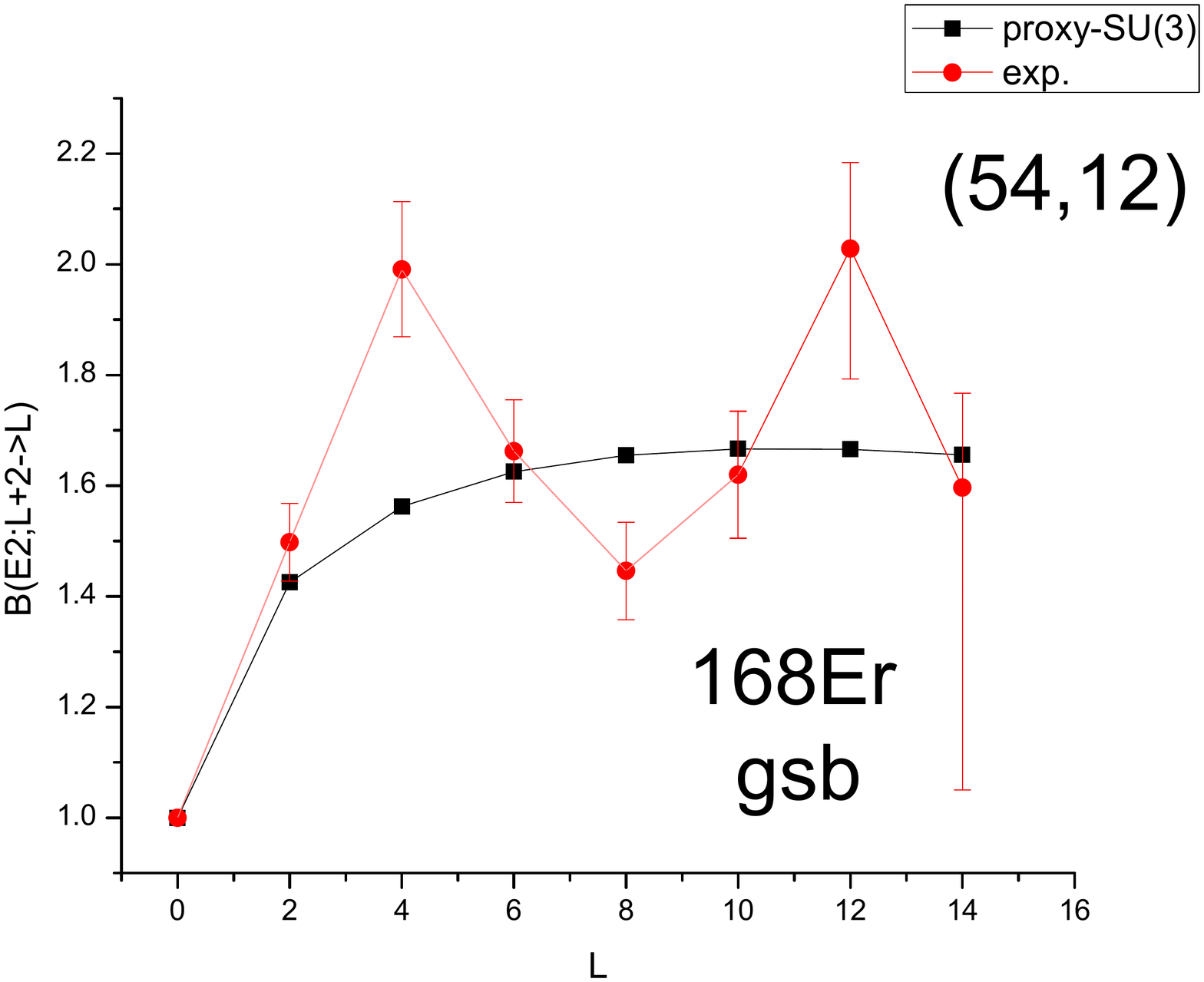,width=55mm}

\epsfig{file=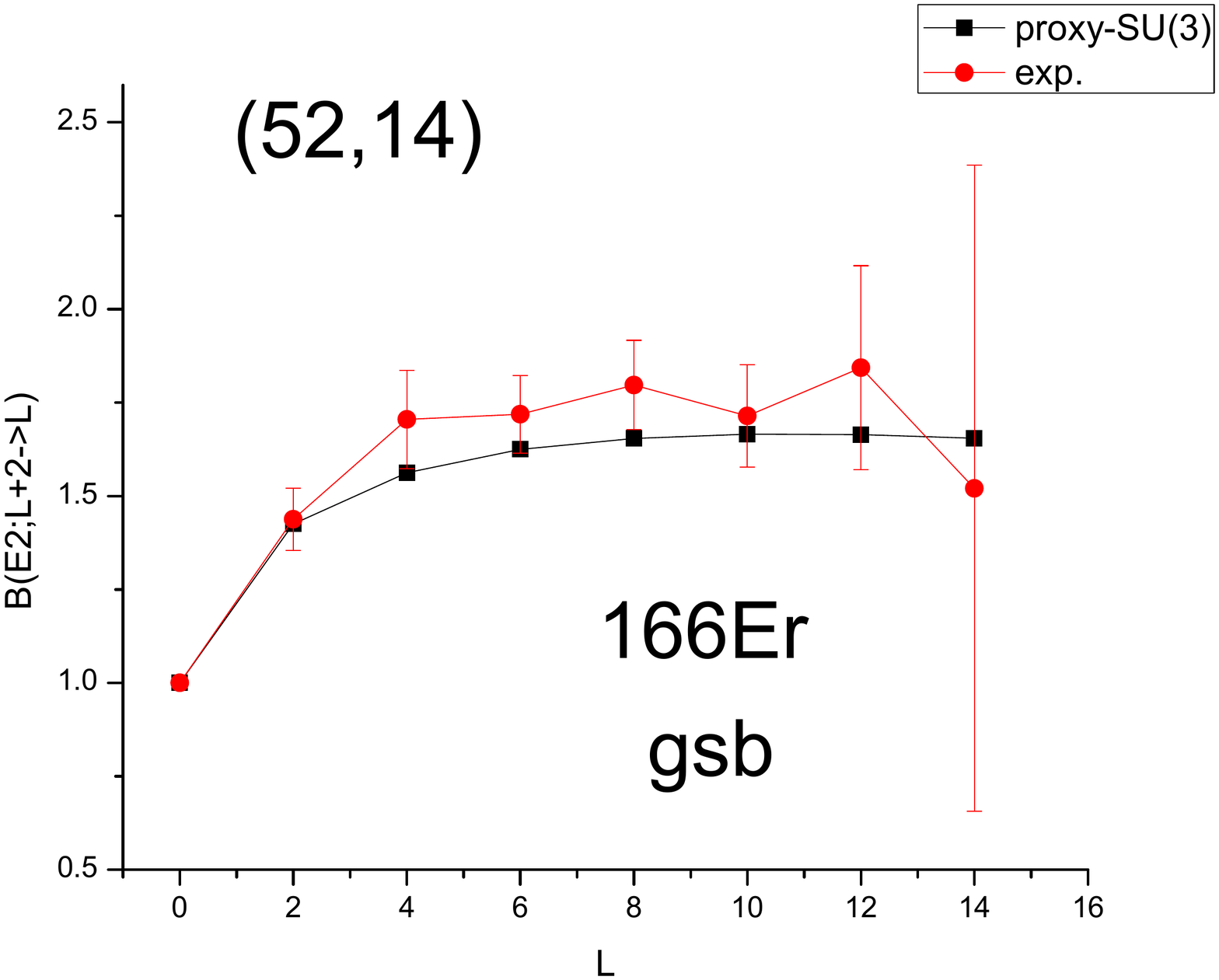,width=55mm}

\caption{B(E2)s within the ground state band are shown for the indicated proxy-SU(3) irreps and for four nuclei, with data taken from \cite{ENSDF}. All values are normalized to $B(E2; 2_1^+ \to 0_1^+)$. 
Results for (54,12) are not shown in the upper left panel, because for this band they are very similar to those of (52,14). See subsection \ref{Exp}  for further discussion.} 
\label{F1}
\end{figure}


\begin{figure}[b]
\epsfig{file=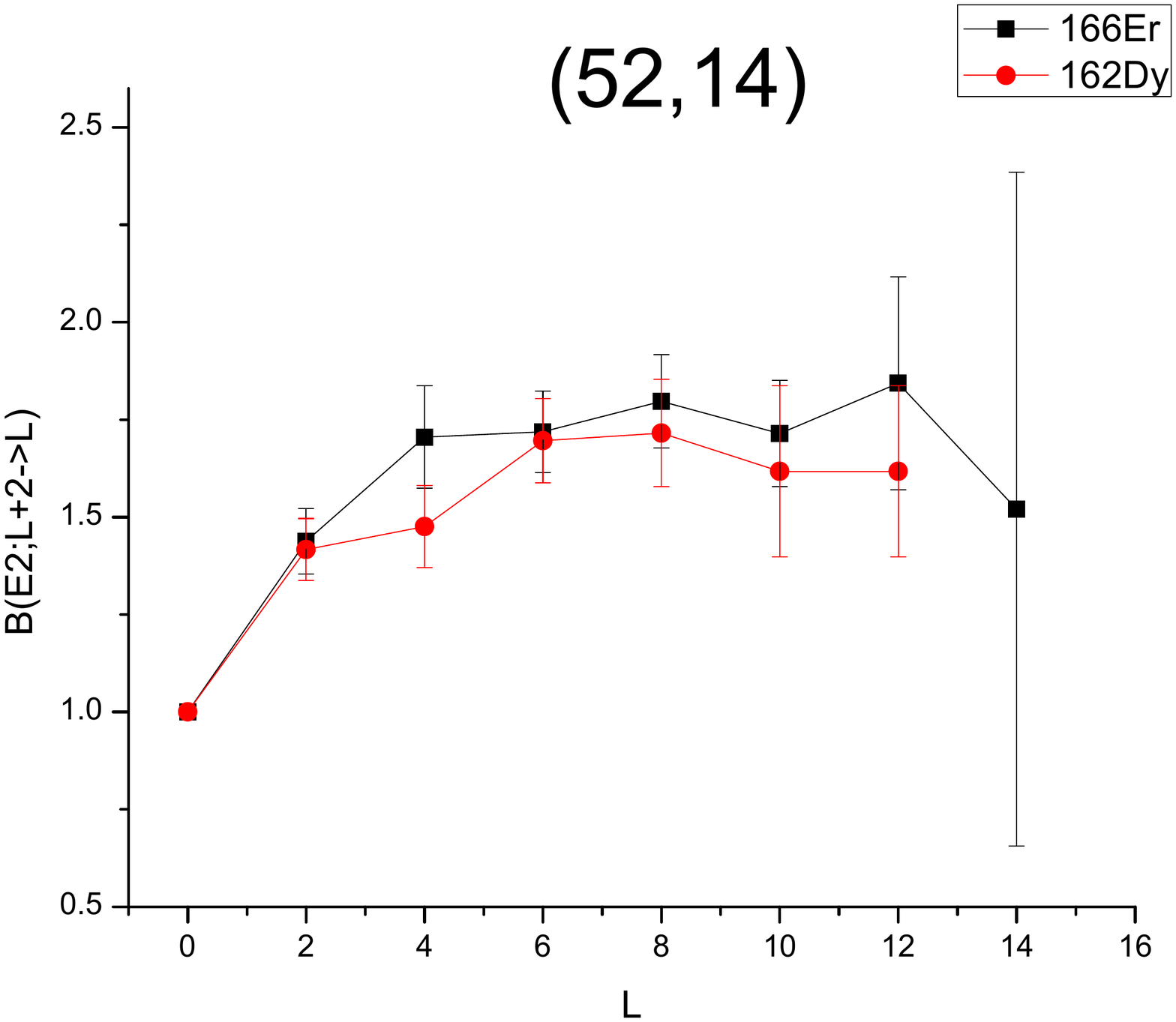,width=55mm}
\epsfig{file=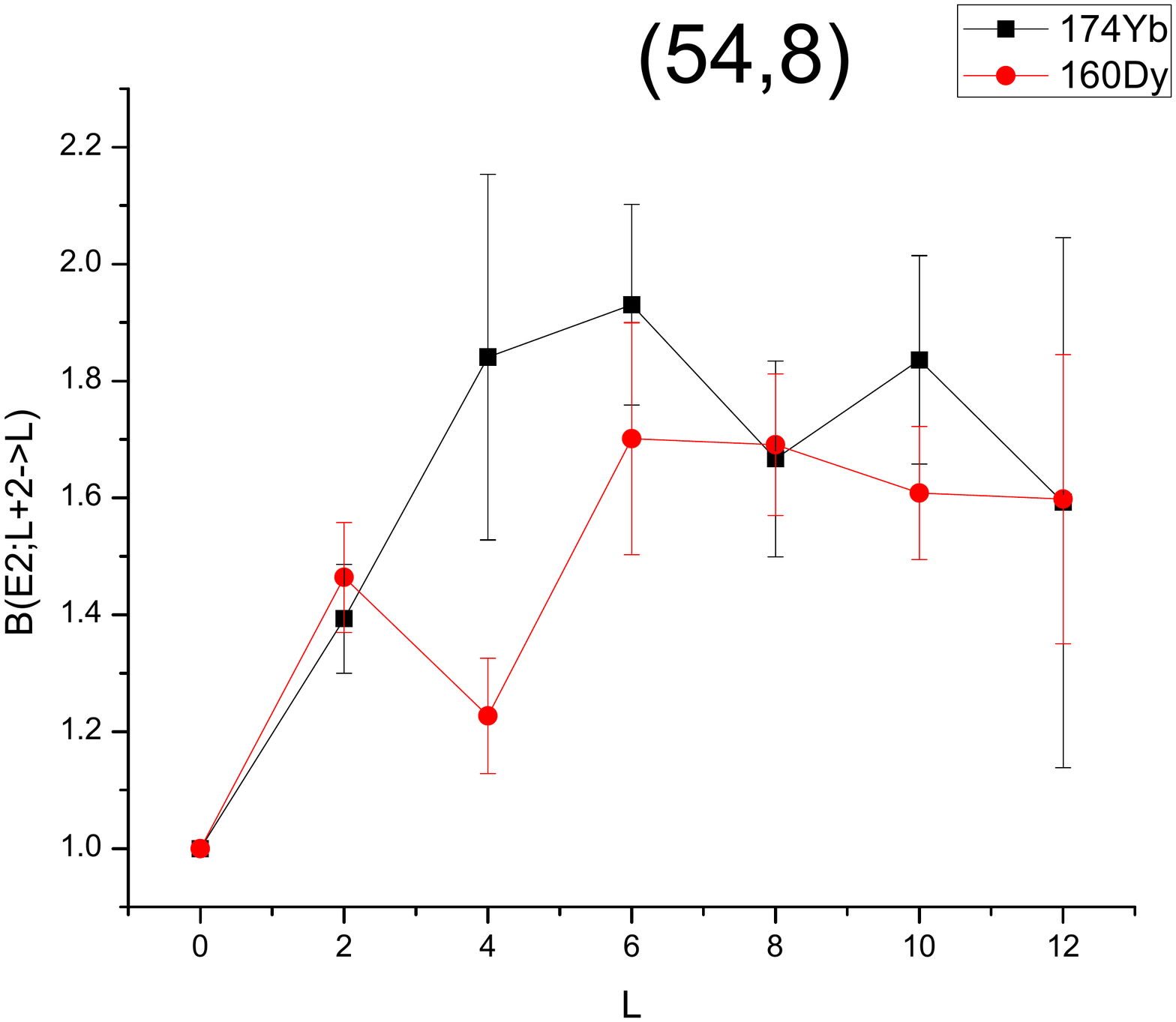,width=55mm}

\epsfig{file=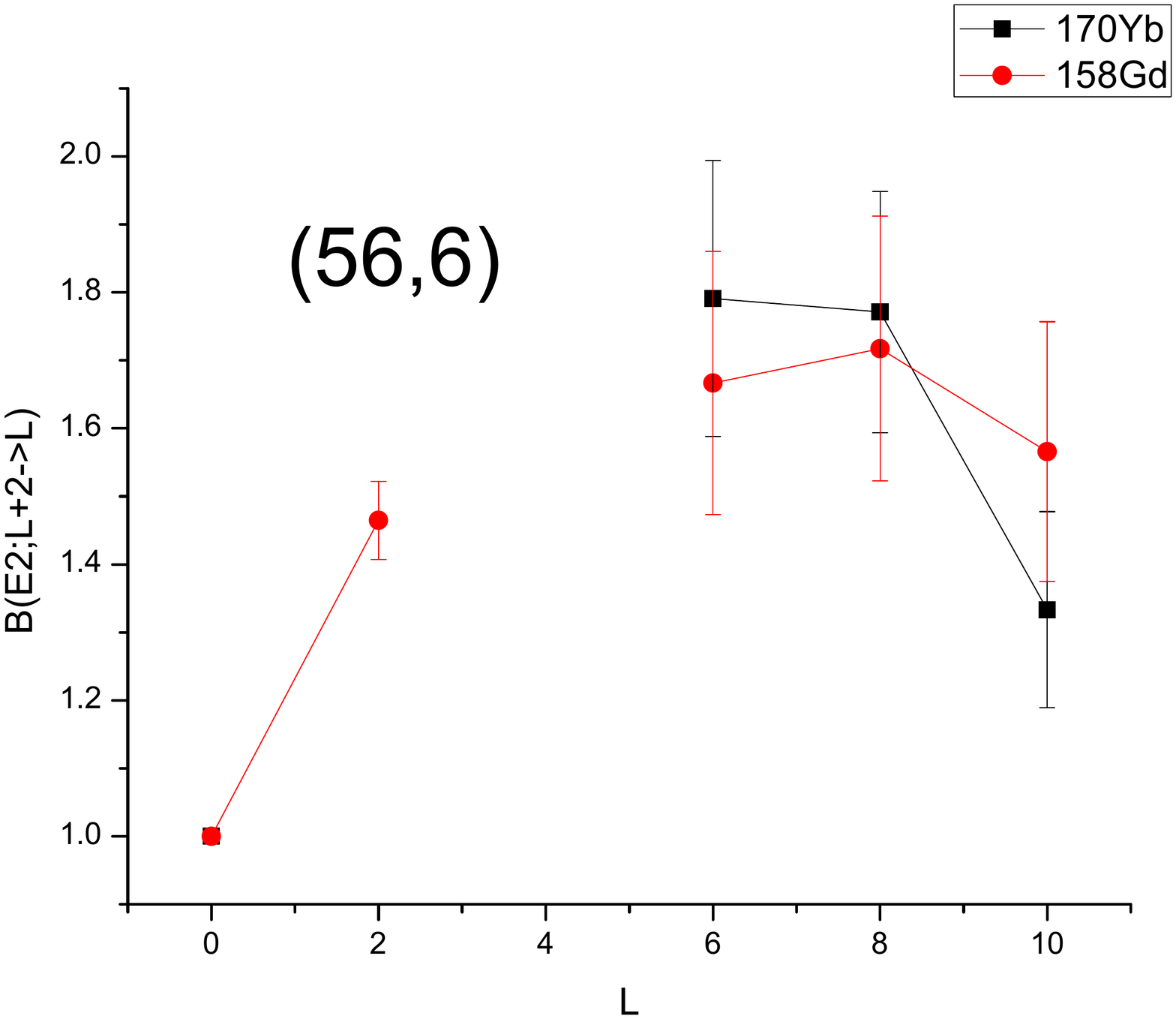,width=55mm}

\caption{Experimental values of B(E2)s within the ground state bands of three pairs of nuclei, each pair accommodated within the same proxy-SU(3) irrep.  Data are taken from \cite{ENSDF}. All values are normalized to $B(E2; 2_1^+ \to 0_1^+)$. 
See subsection \ref{Exp}  for further discussion.} 
\label{F2}
\end{figure}


\begin{figure}[b]

\epsfig{file=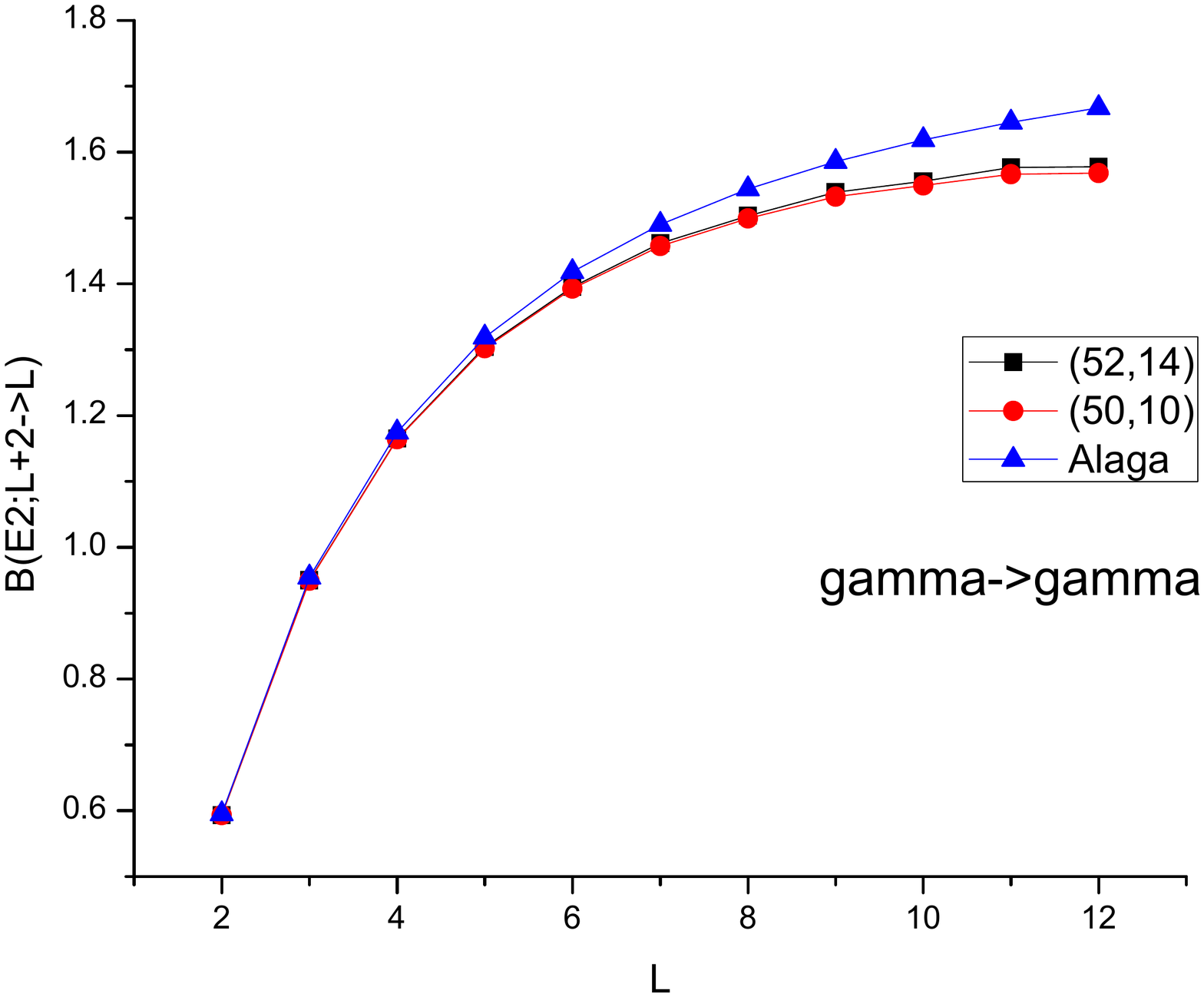,width=55mm}
\epsfig{file=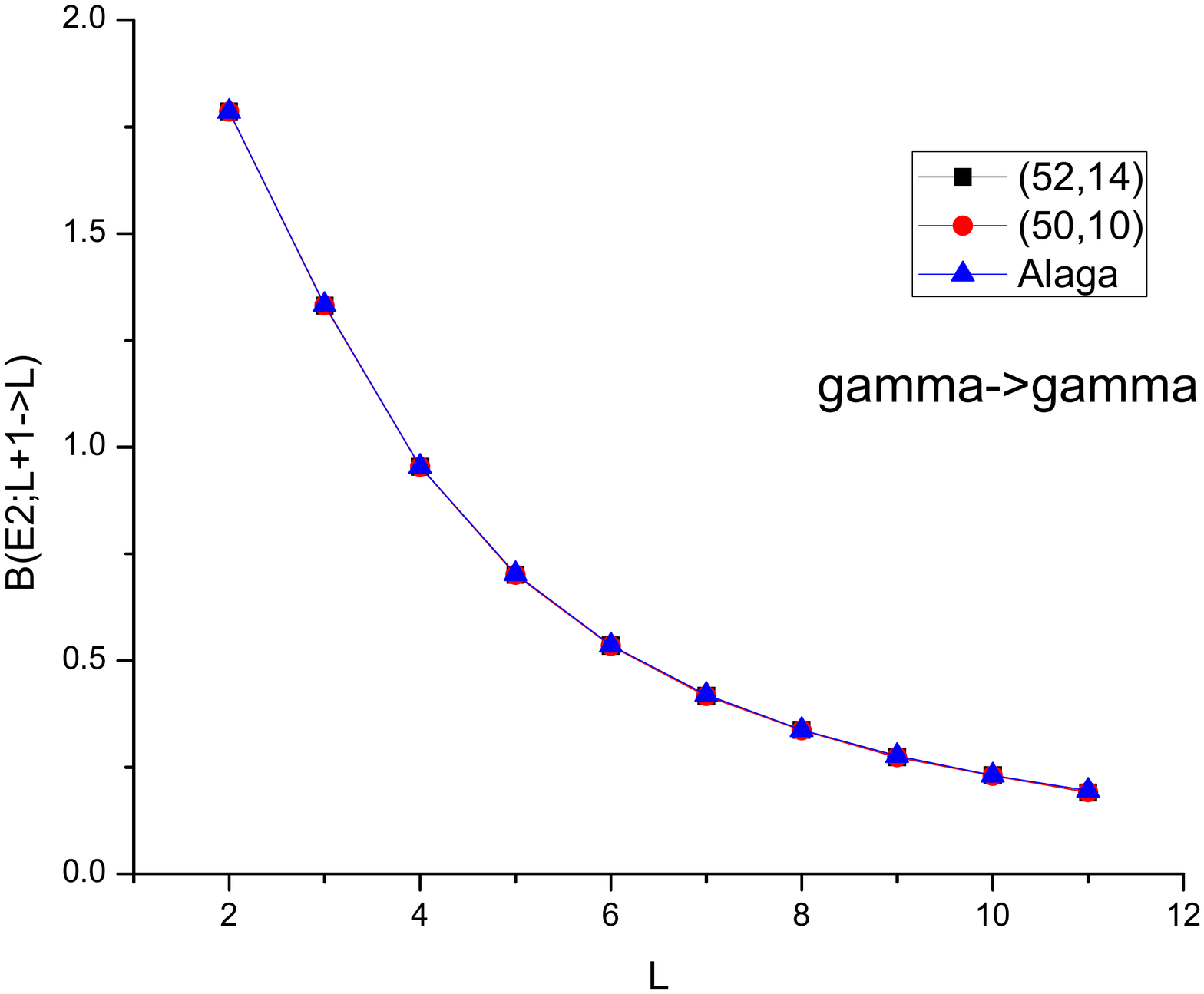,width=55mm}

\epsfig{file=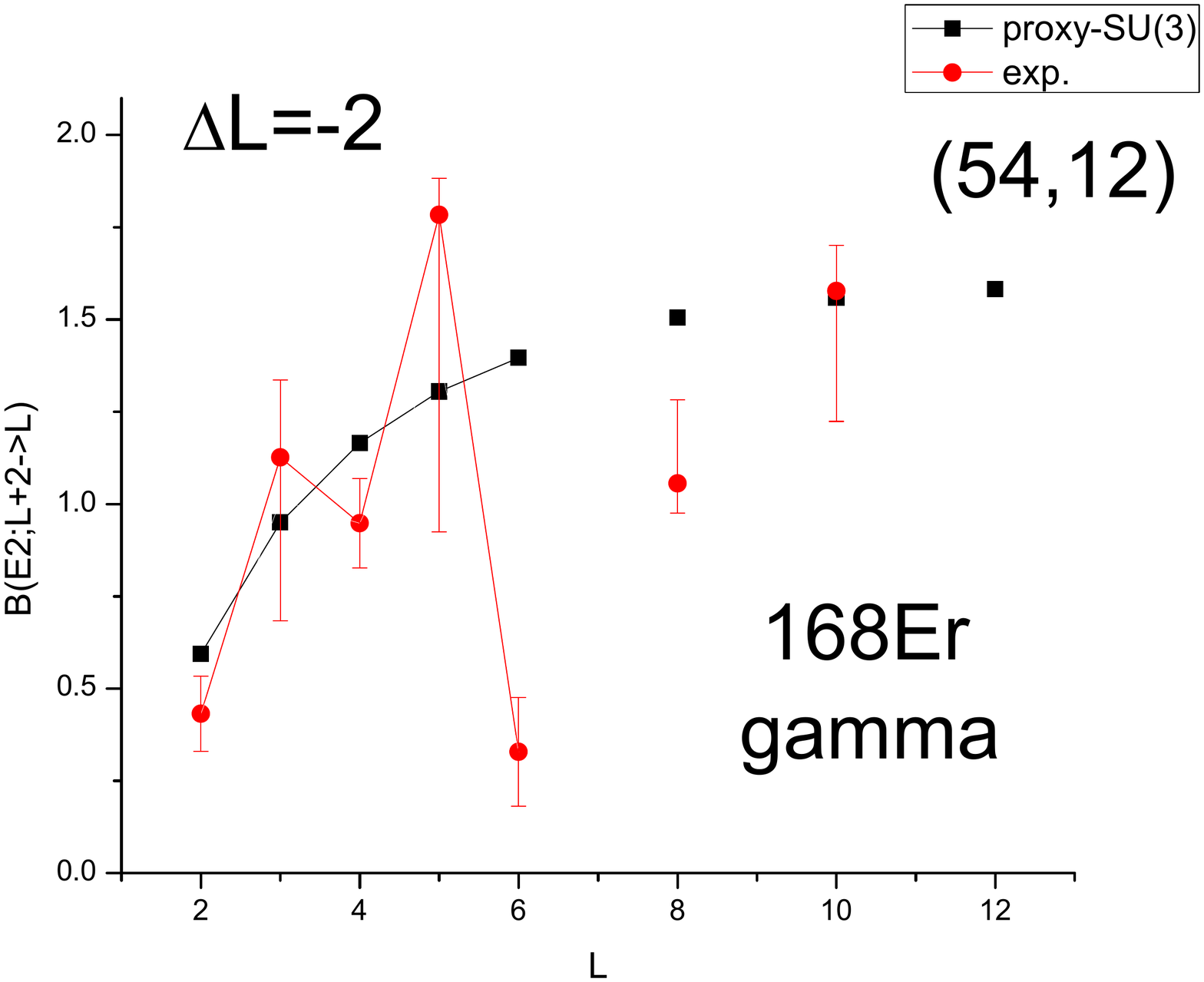,width=55mm}
\epsfig{file=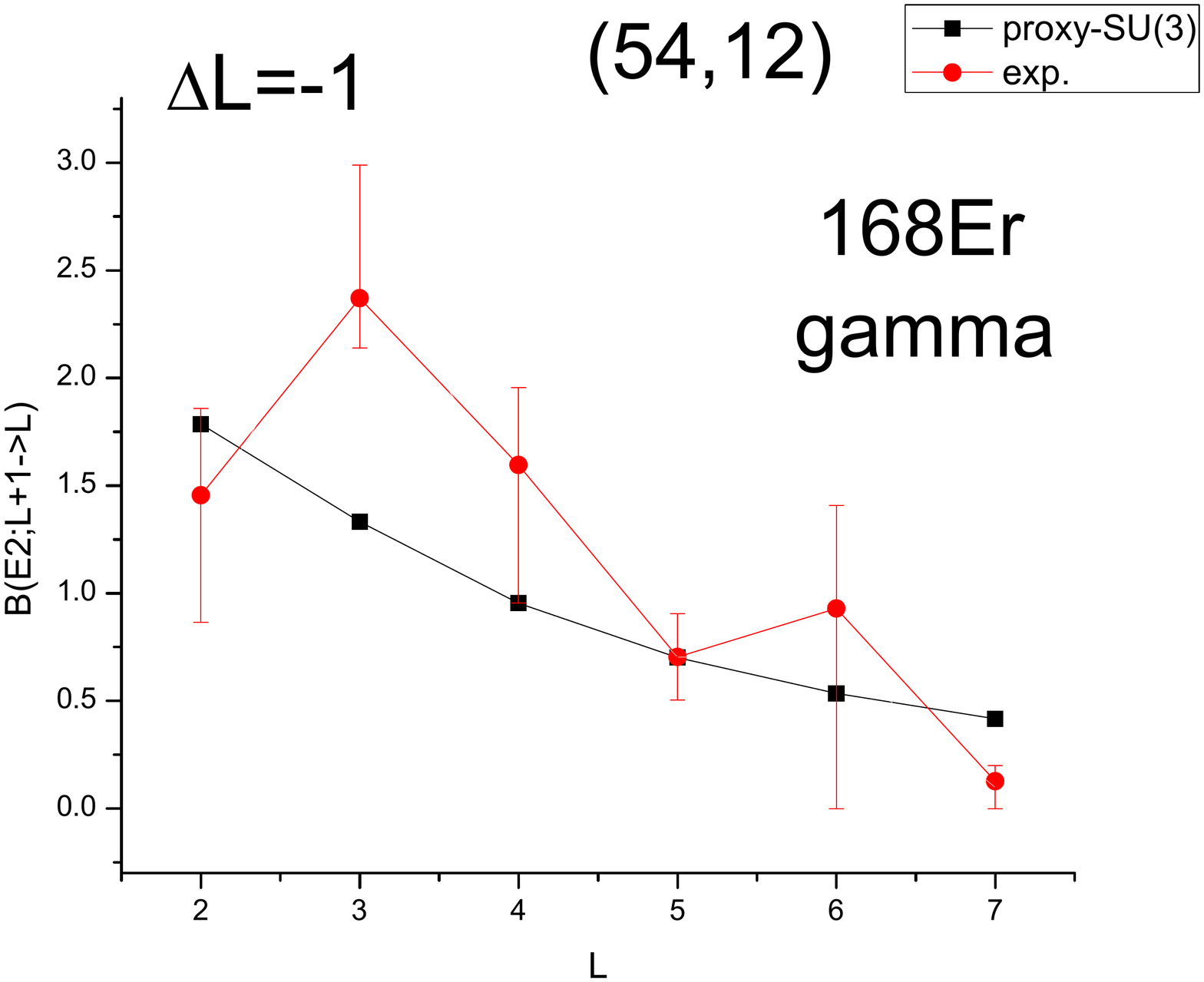,width=55mm}

\epsfig{file=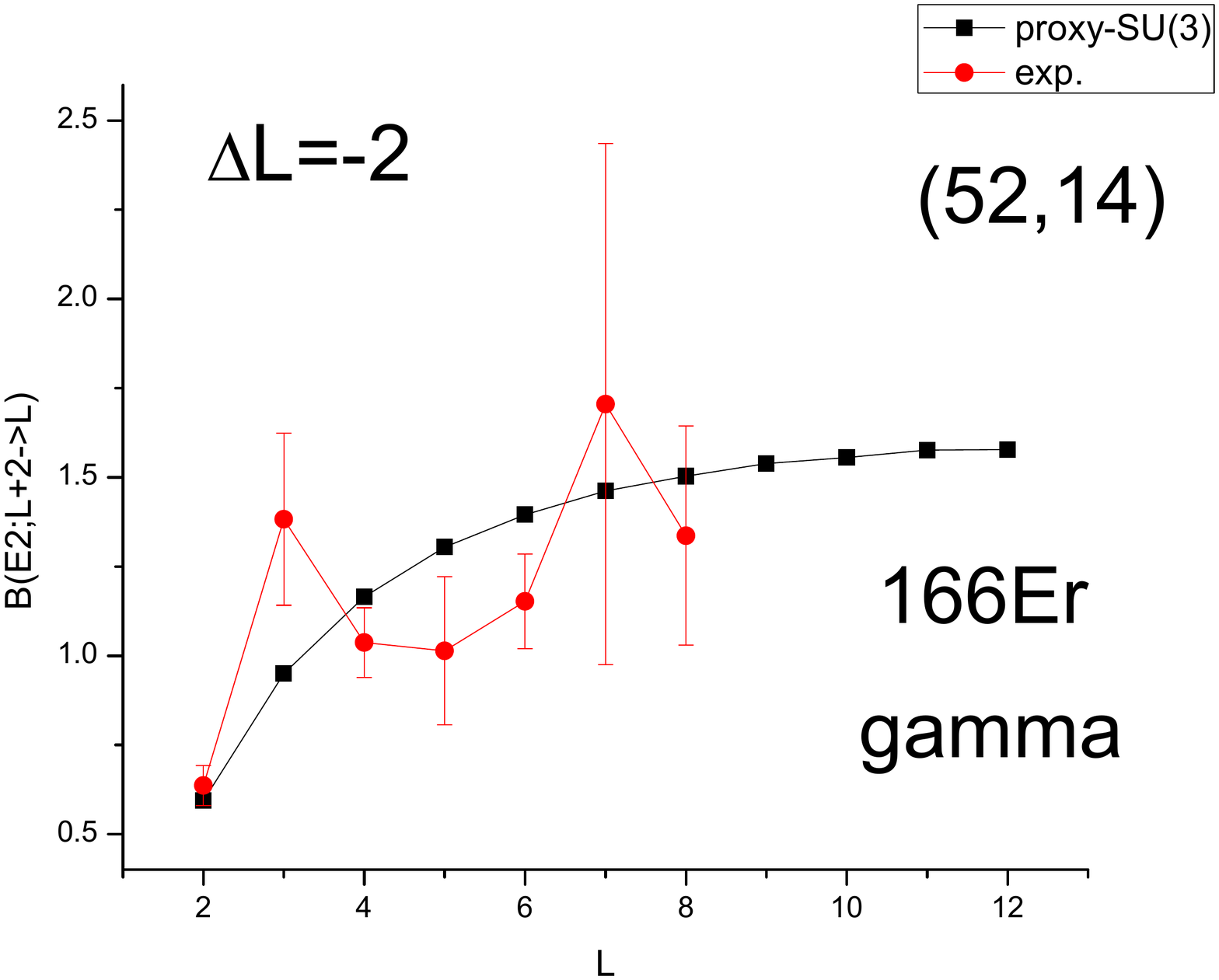,width=55mm}

\caption{B(E2)s within the $\gamma_1$ band are shown for the indicated proxy-SU(3) irreps and for two nuclei, for which sufficient data exist \cite{ENSDF}. All values are normalized to $B(E2; 2_1^+ \to 0_1^+)$. See subsection \ref{Exp}  for further discussion.} 
\label{F3}
\end{figure}

\end{document}